\documentclass[reprint,superscriptaddress,amsmath,amssymb,aps,prx]{revtex4-2}

\usepackage[T1]{fontenc} %
\usepackage[utf8]{inputenc} %
\usepackage{graphicx} %
\usepackage{hyperref}
\usepackage[]{cleveref}

\Crefname{equation}{Eq.}{Eqs.}

\newcommand{\bea}{\begin{eqnarray}}
\newcommand{\eea}{\end{eqnarray}}
\newcommand{\be}{\begin{equation}}
\newcommand{\ee}{\end{equation}}
\newcommand{\ba}{\begin{align}}
\newcommand{\ea}{\end{align}}

\newcommand{\lp}{\left(}
\newcommand{\rp}{\right)}
\newcommand{\lbk}{\left[}
\newcommand{\rbk}{\right]}
\newcommand{\di}{\mathrm{d}}

\newcommand{\T}[1]{\text{#1}}

\newcommand{\sbf}{\boldsymbol{s}}

\begin{document}
	
\title{Cell lineage statistics with incomplete population trees}

\author{Arthur Genthon}
\email{arthur.genthon@hotmail.fr}
\affiliation{Max Planck Institute for the Physics of Complex Systems, 01187 Dresden, Germany}
\affiliation{Gulliver UMR CNRS 7083, ESPCI Paris, Université PSL, 75005 Paris, France}
\author{Takashi Nozoe}
\affiliation{Department of Basic Science, Graduate School of Arts and Sciences, The University of Tokyo, Tokyo 153-8902, Japan}
\affiliation{Research Center for Complex Systems Biology, The University of Tokyo, Tokyo 153-8902, Japan}
\affiliation{Universal Biology Institute, The University of Tokyo, Tokyo 113-0033, Japan}
\author{Luca Peliti} 
\affiliation{Santa Marinella Research Institute, 00052 Santa Marinella, Italy}
\author{David Lacoste}
\affiliation{Gulliver UMR CNRS 7083, ESPCI Paris, Université PSL, 75005 Paris, France}

\begin{abstract}
Cell lineage statistics is a powerful tool for inferring cellular parameters, such as division rate, death rate or the population growth rate. Yet, in practice such an analysis suffers from a basic problem: how should we treat incomplete lineages that do not survive until the end of the experiment? Here, we develop a model-independent theoretical framework to address this issue. 
We show how to quantify fitness landscape, survivor bias and selection for arbitrary cell traits from cell lineage statistics in the presence of death, and we test this method using an experimental data set in which a cell population is exposed to a drug that kills a large fraction of the population. This analysis reveals that failing to properly account for dead lineages can lead to misleading fitness estimations. For simple trait dynamics, we prove and illustrate numerically that the fitness landscape and the survivor bias can in addition be used for the non-parametric estimation of the division and death rates, using only lineage histories. 
Our framework provides universal bounds on the population growth rate, and a fluctuation-response relation which quantifies the change in population growth rate due to the variability in death rate. 
Further, in the context of cell size control, we obtain generalizations of Powell's relation that link the distributions of generation times with the population growth rate, and show that the survivor bias can sometimes conceal the adder property, namely the constant increment of volume between birth and division.
\end{abstract}

\maketitle

\section{Introduction}

Cells are the fundamental object of biology. Sources of cellular heterogeneity are typically blurred in bulk measurements and can only be detected with single-cell analysis. Thanks to advances in experimental techniques (microfluidics, image analyses, sequencing\dots), single-cell measurements are now widely used for phenotype investigations \cite{balaban_bacterial_2004,bakshi_tracking_2021}. However, single-cell measurements when performed in vivo often take the form of snapshots, which means that important dynamical information contained in lineage trees of cell populations (or equivalently population trees) is often lost or hidden \cite{keil_long-term_2017}.

In fact, even when temporal information about lineages is available, as with time-lapse video microscopy, some practical issues remain. One of them concerns lineages that terminate before the end of the experiment. Lineages can end for various reasons: cells can just stop dividing because of changes in the environment, they can die \cite{Biselli_2020,eisenhoffer_crowding_2012,marinari_live-cell_2012}, or be flushed away as a result of dilution \cite{hashimoto_noise-driven_2016,koldaeva_population_2022}. What should be done in practice with these dead lineages: should they simply be discarded or should they be 
included in the analysis and how?
Clearly, both options are not equivalent and a wrong methodology could lead to unwanted statistical biases. 
In this paper, we propose a theoretical framework to address this issue. Our approach is not limited to cell lineages, it could also be applied to other types of lineage trees containing extinct lineages, which appear for instance in evolution experiments. 

A key idea in lineage analysis is the distinction between the backward sampling which reflects the differences in reproductive success between cells, and the forward sampling that balances these differences \cite{nozoe_inferring_2017,yamauchi_unified_2022}. We first review this formalism in \cref{sec_background}. Then, we generalize this approach to include dead lineages in \cref{sec_back_for}. This framework can be used with any given lineage tree independently of the dynamics that generate the tree and of the cause of early lineage ending, and it is valid at arbitrary finite time. 

Based on this framework, in \cref{sec_fit} we propose measures of fitness landscape and survivor bias, particularly adapted to cases where the cell traits of interest are correlated with the cause of death. We test these ideas using experimental data of mycobacteria subject to antibiotics \cite{wakamoto_dynamic_2013}, an analysis which reveals the importance of correctly accounting for dead lineages.
These measures of fitness and survival are very general and can be evaluated for any lineage tree. For some simple models of population dynamics, we show that they can also be used to infer the division and death rates using phenotypic trajectories.
These measures are further used to theoretically quantify selection in the presence of death and fitness variability, and the death-induced change in selection.

In \cref{sec_thermo}, we explain the analogy between our framework and that of stochastic thermodynamics \cite{seifert_stochastic_2012}.
From this analysis, we deduce universal bounds on the population growth rate in the presence of death, which we test with experimental data, and we derive a fluctuation-response relation which relates the change in population growth rate to the variability in death rate. 

Finally, we apply our formalism to models of cell size control in \cref{sec_csc}, where we derive modified versions of Powell's relation \cite{powell_growth_1956} in the presence of death. These relations link the backward and forward distributions of inter-division times and imply that cells divide faster under selection. A consequence of death on the statistics of the adder mechanism of cell size control
\cite{taheri-araghi_cell-size_2015}, also follows from this approach.

\section{Theoretical background for complete lineage trees}
\label{sec_background}

\begin{figure*}
	\includegraphics[width=0.8\linewidth]{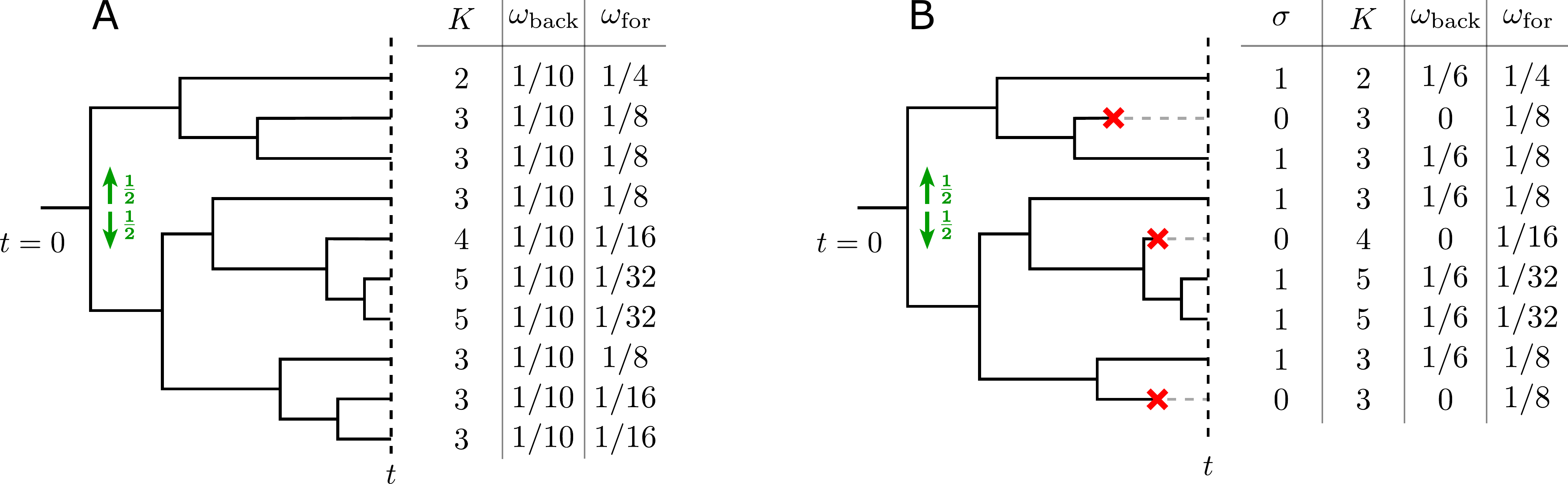}
	\caption{Population trees without and with death. (A): Population tree with no death starting with $N_0=1$ cell and ending with $N(t)=10$ lineages at time $t$. In the table, the columns indicate the number $K$ of divisions along the lineage and the backward and forward weights. 
	(B): Population tree starting with $N_0=1$ cell and ending with $n(\sigma=0,t)=3$ dead lineages (red crosses), and $N(t)=n(\sigma=1,t)=6$ alive lineages at time $t$. In the table, the survival status $\sigma$ takes value $1$ for surviving lineages and $0$ for dead lineages. 
	The forward probability of survival is computed as:
		$p_{\T{for}} (\sigma=1,t)=1\times 1/4 + 3\times 1/8 + 2\times 1/32=11/16$.}
	\label{fig_trees}
\end{figure*}

In this introductory section, we review the framework proposed in \cite{nozoe_inferring_2017,yamauchi_unified_2022} to analyze lineage trees in the absence of cell death. 

From this framework, it is possible to infer fitness differences associated with distinct states of cellular trait
and selection within a growing population from empirical lineage tree data.
For instance, time-lapse single-cell measurements provide cellular divisions and trait time series
(e.g. fluorescence intensity of reporter protein) in the form of lineage trees.
Because phenotypic traits temporally fluctuate, time series branch at division in such datasets,
and it is therefore not trivial what statistics correctly report fitness value of phenotypic trait.

To resolve this issue, we assign two types of probability weight to each cellular lineage $l$:
the backward weight $\omega_\mathrm{back}(l) = N(t)^{-1}$ and the forward weight $\omega_\mathrm{for}(l) = N_0^{-1} m^{-K(l)}$, where $N_0$ and $N(t)$ are the numbers of cells at initial time $t=0$ and final time $t$, $K(l)$ is the number of divisions along lineage $l$, and $m$ is the constant 
number of daughter cells born at division (including the mother cell). An example of lineage tree with the corresponding forward and backward weights is represented in \cref{fig_trees}A.
The backward sampling weighs uniformly the cells present at final time, thus leading to an over-representation of cells with above-average reproductive success. Instead, the forward weight represents the probability of choosing a cell lineage descending the tree from one of the ancestor cells at time $t=0$ and selecting one branch among the $m$ possibilities with equal probability $1/m$ at each division. Doing so, a lineage with high reproductive success is sampled with a lower weight in the forward sampling than in the backward sampling, in such a way as to balance the selection effect.
Therefore, the comparison between these two samplings informs on the selection undergone by lineages, as detailed below.

Let us now consider a general cell trait $\mathcal{S}$, taking values $s$, which can for example represent a phenotype or a genotype. This trait can be any property of a lineage: a snapshot property evaluated at a given time, an averaged trait along lineages or a trait trajectory, for instance. 
The discussion above motivates us to compute the forward (backward) probability to pick a lineage with
$K$ divisions and with trait value $s$ in forward (backward) manner.
Let $n\left(K,s,t\right)$ denote the number of such lineages, we then define
\begin{align}
	\label{eq_def_p_back_circ}
	p_{\rm{back}}(K,s,t) & =  N(t)^{-1} n(K,s,t)  \\
	\label{eq_def_p_for_circ}
	p_{\rm{for}}(K,s,t)&=N_0^{-1} m^{-K} n(K,s,t)\,.
\end{align}

The fitness of the exponentially growing population is defined as its growth rate:
\begin{equation}
\label{eq_def_lambda}
\Lambda_t = \frac{1}{t}\ln \frac{N(t)}{N_0} \,,
\end{equation}
and using the forward and backward distributions, we obtain another representation of population fitness:
\begin{equation}
	\Lambda_t = \frac{1}{t}\ln \left\langle m^{K} \right\rangle_\mathrm{for} \,,
	\label{eq:popfitness_nodeath}
\end{equation}
where $\left\langle\cdot\right\rangle_\mathrm{for}$ denotes the average with respect to $p_\mathrm{for}(K,s,t)$.
In this representation, 
$m^{K}$ and $p_\mathrm{for}\left(K,s,t\right)$ are understood as the fitness value of a lineage and its probability distribution prior to selection.

Since the fitness landscape is conventionally defined as the mean fitness conditioned on a genotype or phenotypic trait,
it is natural to define the fitness landscape of cell trait $\mathcal{S}$ as
\begin{equation}
	h_t(s) = \frac{1}{t} \ln \left\langle m^K \vert s\right\rangle_\mathrm{for} \,,
	\label{eq:fitnesslandscape_nodeath}
\end{equation}
where $\left\langle\cdot\vert s \right\rangle_\mathrm{for}$ denotes the conditional average with respect to $p_\mathrm{for}(K,t|s)=p_\mathrm{for}(K,s,t)/p_\mathrm{for}(s,t)$,
with $p_\mathrm{for}\left(s,t\right) = \sum_{K} p_\mathrm{for}\left(K,s,t\right)$.
Using \cref{eq_def_p_back_circ,eq_def_p_for_circ}, together with $p_\mathrm{back}\left(s,t\right) = \sum_{K} p_\mathrm{back}\left(K,s,t\right)$, the fitness landscape of cell trait $\mathcal{S}$ is rewritten as
\begin{equation}
	h_t(s) = \Lambda_t + \frac{1}{t} \ln \frac{p_\mathrm{back}\left(s,t\right)}{p_\mathrm{for}\left(s,t\right)} \,.
	\label{eq:fitnesslandscape2_nodeath}
\end{equation}
In this alternative form, the difference $h_t(s)-\Lambda_t$ between the fitness value of a cell with trait $s$ and the population fitness measures  the over- or under-representation of lineages with this trait value $s$ 
in the growing population as compared to the statistics prior to selection. 

To quantitatively evaluate the overall effect of selection on the trait statistics, we define the strength of selection as the gain in mean fitness when ‘turning on' selection by changing the distribution of trait $\mathcal{S}$ from the forward ensemble, where differences in reproductive success are suppressed by the weighting of the lineages, to the backward ensemble:
\begin{equation}
	\Pi_{\mathcal{S}} = \left\langle h_t \right\rangle_\mathrm{back} - \left\langle h_t \right\rangle_\mathrm{for} \,.
	\label{eq:selstr_nodeath}
\end{equation}
Because $\Pi_{\mathcal{S}}$ is positively proportional to the Jeffrey divergence \cite{nozoe_inferring_2017}, a non-negative symmetrical information-theoretic distance between the forward and backward distributions, it is always non-negative. 
If the fitness landscape is used to determine which values $s$ of a given cell trait $\mathcal{S}$ are selected in a growing population,  on the other hand the strength of selection is a measure of the overall selection acting on trait $\mathcal{S}$, and can thus be compared to other cell traits $\mathcal{S'}$ to determine which traits are under the strongest selection, i.e. which traits strongly correlate with lineage fitness.

Importantly, we want to emphasize that the notions of fitness and selection introduced here are different from their usual definitions based on division/growth rates. Usually, fitness quantifies the chance to reproduce, and thus measures the amplification over time of the proportion of some cell traits in the population. However, this measure does not account for the already-existing distribution of cell traits prior to selection.
The fitness landscape proposed above instead measures the amplification of the forward (or a priori) probabilities when cells are competing in a population. 

\section{A framework to sample incomplete lineage trees}
\label{sec_back_for}

\begin{figure}
    \includegraphics[width=\linewidth]{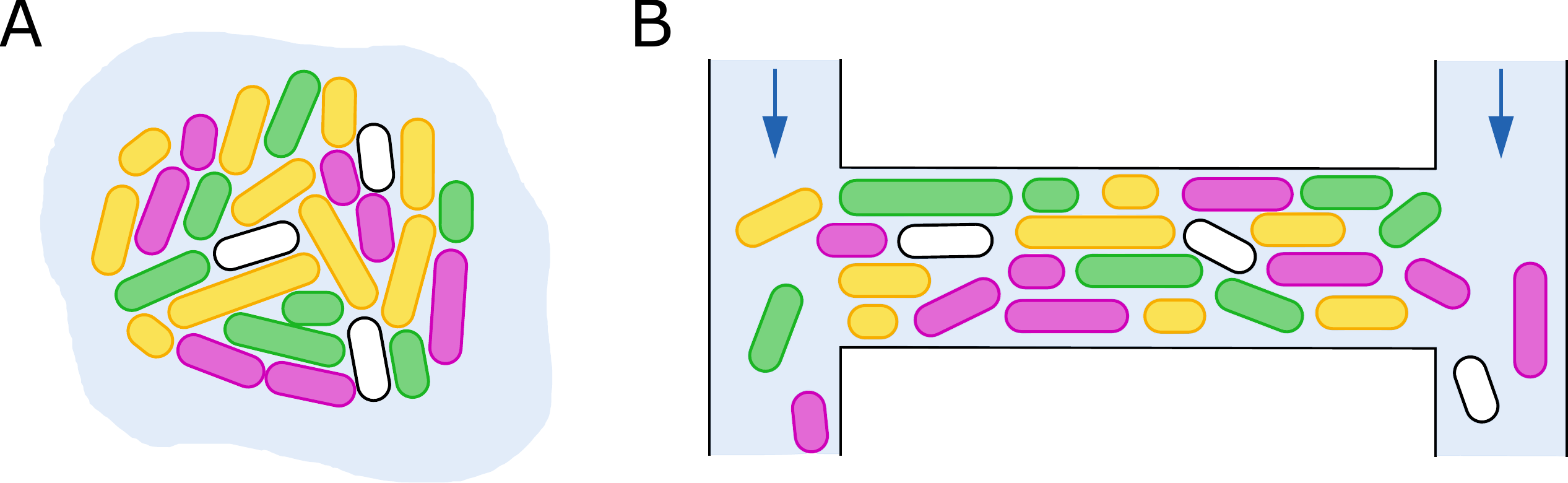}
\caption{Cartoons of two experimental setups where lineages can end before the end of the experiment. The cell colors represent cell heterogeneity and the white cells are dead. (A): free growth in bulk; cells may be far from each other or in contact as in a tissue. (B): cytometer setup from \cite{hashimoto_noise-driven_2016} where cells grow in a chamber open at both ends, and are evacuated by the flow of growth medium in order to maintain the population constant inside the chamber.}
	\label{fig_setups}
\end{figure}

The main experimental setups used to illustrate our framework are represented on \cref{fig_setups}: (A): free growth and (B) growth in a micro-channel \cite{hashimoto_noise-driven_2016}. The cell colors are an illustration of the bacterial diversity, and white cells are dead. 
In the micro-channel, the down arrow indicates the medium flux that carries away cells that go out of the channel, so that both biological death and dilution can be present, while no dilution is observed in free growth. The case of free growth does not require cells to be far from each other, they can be strongly coupled or in contact as it happens in tissues for instance. Both cases A and B are represented by a population tree like the one shown in \cref{fig_trees}B, where the red crosses indicate death in a broad meaning. Such a tree starts with $N_0$ cells at time $t=0$ and ends with $N(t)$ living cells at time $t$. Lineages can either survive up to time $t$ or die before, as indicated by $\sigma$, taking value $1$ for alive lineages and $0$ for dead lineages. Note that $\sigma$ refers to the status (dead or alive) of lineages at the final time $t$, irrespective of the time of death.

When introducing death, the two samplings introduced in \cref{sec_background} are modified in the following way. 
When taking a snapshot of the population at time $t$, only living lineages are present, and in the backward sampling we sample them uniformly with weights
$\omega_{\T{back}}(l)=N(t)^{-1} \delta(\sigma(l)-1)$, where $\delta(x)$ is a Kronecker delta. 
On the other hand, starting from $t=0$ with one of the $N_0$ initial cells and following the lineages up to time $t$ by choosing with uniform probability $1/m$ one of the $m$ daughter cell born at each division, both dead and living lineages are sampled with the forward weights 
$\omega_\T{for}(l)= N_0^{-1} m^{-K(l)}$,
where $K(l)$ is the number of divisions along lineage $l$ up to time $t$. We give a simple example of how these weights are computed in practice in \cref{fig_trees}B.
A major difference with the deathless case immediately appears: some lineages are sampled in the forward manner but not in the backward manner. 

For a general cell trait $\mathcal{S}$, the forward and backward probabilities to pick a lineage with $K$ divisions, with survival status $\sigma$ and trait value $s$ are modified accordingly:
\begin{align}
	\label{eq_def_p_back}
	p_{\rm{back}}(K,\sigma,s,t) & =  N(t)^{-1} n(K,\sigma,s,t) \delta(\sigma-1) \\
	\label{eq_def_p_for}
	p_{\rm{for}}(K,\sigma,s,t)&=N_0^{-1} m^{-K} n(K,\sigma,s,t)\,.
\end{align}
where  $n(K,\sigma,s,t)$ is the number of such lineages. The number of cells alive in the population at time $t$ is now given by $N(t)=n(\sigma=1,t)=\sum_{K,s} n(K,s,\sigma=1,t)$.

The forward probability of survival $p_{\T{for}} (\sigma=1,t)=\sum_{K,s} p_{\T{for}} (K,s,\sigma=1,t)$ is a central quantity in this problem. 
It is a strictly decreasing function of the number of death events, and is unaffected by divisions, therefore it tends to $0$ as $t \rightarrow \infty$. Indeed, when a lineage with $K$ divisions, associated with a weight $N_0^{-1} m^{-K}$ in the forward sampling, divides in $m$ daughters cells all weighted $N_0^{-1} m^{-(K+1)}$, then the overall weights of the $m$ daughters is equal to that of the mother, and the forward weight is conserved. On the other hand, when a lineage dies its forward weight disappears from the sum over living lineages that defines $p_{\T{for}} (\sigma=1,t)$, leading to a decrease of the latter.
We then define the rate of decrease of the forward probability of survival as:
\begin{equation}
\label{eq_def_Gamma}
\Gamma_t=\frac{1}{t} \ln p_{\T{for}} (\sigma=1,t) \,.
\end{equation}

Finally, we introduce the forward distribution conditioned on survival:
\begin{equation}
\label{eq_def_p_star}
p^{\star}_{\T{for}}(\cdot,t)=p_{\T{for}}(\cdot,t|\sigma=1) \,.
\end{equation}

\section{Quantifying fitness and selection in incomplete lineage trees}
\label{sec_fit}

\subsection{Disentangling fitness and survival}
\label{sec_fit_surv}

In the absence of death, the measure of fitness defined by \cref{eq:fitnesslandscape_nodeath,eq:fitnesslandscape2_nodeath}, which was introduced in \cite{nozoe_inferring_2017} and reviewed in \cref{sec_background}, depends only on the correlations between the value of the trait and the number of divisions along the lineage.
When introducing death, the frequency of the trait in the population is the result of the correlations between this trait and the divisions, together with the correlations between this trait and survival. 
Therefore, the fitness landscape defined in \cref{eq:fitnesslandscape_nodeath,eq:fitnesslandscape2_nodeath} captures both fitness and survival effects. Ignoring dead lineages and using this measure to quantify selection could thus lead to misleading fitness estimations as we will demonstrate. 

To disentangle the two effects, we define the fitness landscape in the presence of incomplete lineages as the average fitness conditioned on the cell trait for surviving lineages only:
\begin{equation}
	\label{eq_def2_h_death}
	h^{\star}_t(s)=\frac{1}{t}\ln \lbk \sum_K m^K p_{\T{for}}^{\star}(K,t|s) \rbk \,.
\end{equation}
We show in \cref{sec_leibler} that this fitness landscape is consistent with notions of fitness proposed previously. In particular, it recovers the historical fitness introduced in \cite{leibler_individual_2010} for populations undergoing stochastic phenotype switching, while being more general because it is model-independent.
When $p_{\T{for}}^{\star}(K,t|s) = p_{\T{for}}^{\star}(K,t)$ for any $s$, then the fitness landscape is flat and equal to $\Lambda_t-\Gamma_t$. This condition is called the conditional independence of $K$ and $s$ knowing $\sigma=1$. Be careful that the conditional independence does not imply, and is not implied by, the regular independence between $K$ and $s$. This means in particular that a trait $s$ can be correlated to both $K$ and $\sigma$, and still have a flat landscape. 

As in \cref{sec_background}, by expressing $p_{\T{for}}^{\star}(K,t|s)$ in \cref{eq_def2_h_death} and using \cref{eq_def_p_for,eq_def_p_back,eq_def_Gamma,eq_def_p_star}, the fitness landscape can be reshaped as
\begin{equation}
\label{eq_def_h_death}
h^{\star}_t(s)=\Lambda_t - \Gamma_t + \frac{1}{t} \ln \lbk \frac{p_{\T{back}}(s,t)}{p^{\star}_{\T{for}}(s,t)} \rbk \,,
\end{equation}
where the marginalized trait distributions are defined as $p(s,t)=\sum_K p(s,K,t)$ (in both backward and forward). By comparing the forward and backward samplings for surviving lineages only, we get rid of the incomplete lineages and thus isolate selection effects from survival effects. 

When combining the previous definition \cref{eq_def_h_death} with \cref{eq_def_p_for,eq_def_p_back,eq_def_Gamma,eq_def_p_star}, the fitness landscape can also be written as 
\begin{equation}
	\label{eq_def2_h_death_back}
	h^{\star}_t(s)=-\frac{1}{t}\ln \lbk \sum_K m^{-K} p_{\T{back}}(K,t|s) \rbk \,.
\end{equation}
This formulation has a form similar to \cref{eq_def2_h_death}, but measures the inverse linage fitness $m^{-K}$ averaged with respect to the probability distribution conditioned on the cell trait and posterior to selection.

We finally define the survivor bias as:
\be
\label{def_h_dag}
h_t^{\dagger}(s) = h_t(s) - h^{\star}_t(s) =\Gamma_t + \frac{1}{t} \ln \lbk \frac{p^{\star}_{\rm{for}} (s,t)}{p_{\rm{for}} (s,t)} \rbk \,.
\ee
which measures the statistical difference between the ensemble of all lineages and the surviving lineages only. 
While this definition of survivor bias is not the only possible one, it is reasonable to compare the two lineage ensembles using the forward sampling, in order to get rid of selection effects and isolate survival effects. 
This bias is significant when considering a trait $\mathcal{S}$ which is correlated with survival. One important example is antibiotic resistance, because there, one is interested precisely in traits which appear among surviving bacteria in the presence of antibiotics \cite{wakamoto_dynamic_2013}. We explore this example in the next section.

\subsection{Making sense of antibiotics experiments}

\begin{figure*}
	\includegraphics[width=\linewidth]{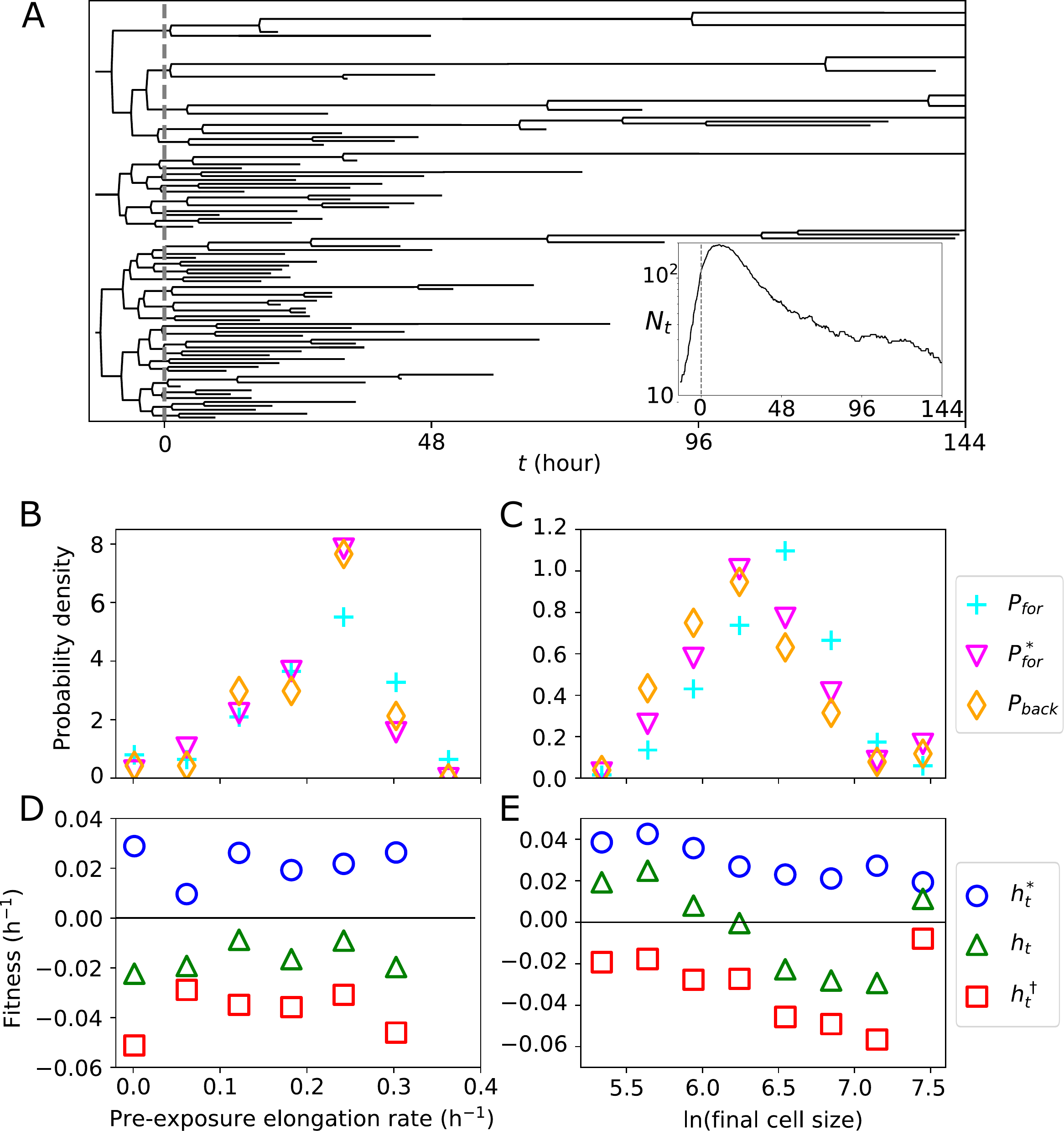}
	\caption{Analysis of a lineage tree with biological death.
		(A) Example of lineage tree of individual bacterial cells proliferating and being killed by drug exposure \cite{wakamoto_dynamic_2013}.
		The population is exposed to INH on a microfluidic device from $t=0$ h, as indicated by the dashed line.
		Inset: evolution of the corresponding number $N_t$ of alive cells with time.
		(B) and (D): Distributions and landscapes of the pre-exposure growth rate (computed one hour before drug exposure). There are no values at the rightest bin for $h^{\star}_t$, $h_t$ and $h^{\dagger}_t$ because there are no alive lineages with this value of pre-exposure growth rate ($p_{\rm{for}}^{\star}=p_{\rm{back}}=0$).
		(C) and (E): Distributions and landscapes of the logarithmic cell size at the end of the lineage (computed at $t=36$ h for surviving lineages or at the time of death for lineages that died before). 
		}
\label{fig_tree_landscape_death}
\end{figure*}

To illustrate the importance of the distinction between $h^{\star}_t$, $h_t$ and $h^{\dagger}_t$ when analyzing experimental data, we use lineage trees of
{\it Mycobacterium smegmatis} exposed to the drug isoniazid (INH) \cite{wakamoto_dynamic_2013}. 
After a few divisions in the absence of the drug, the population is exposed to the drug over 6 days and every individual cell lineage is tracked until it is killed or reaches the end of observation (\cref{fig_tree_landscape_death}A). 
In the original paper \cite{wakamoto_dynamic_2013}, the authors tested the hypothesis that bacterial persistence, i.e. the survival of a small fraction of cells when exposed to antibiotics which is not linked to a genetically-acquired resistance, was linked to the existence of persisters that grow very slowly or do not grow at all before drug exposure, as observed for \textit{E. coli} in \cite{balaban_bacterial_2004}. To do so they compared the pre-exposure elongation rates of persistent vs non-persistent cells, and observed no significant difference. They thus concluded that the pre-exposure elongation rate is not correlated with survival for {\it M. smegmatis}. 

We first aim to reproduce this observation with a formalism adapted to populations with dead lineages, and to further study the correlation between this cell trait and selection. In addition, we exhibit another cell trait that is correlated with selection and survival: the logarithmic cell size at the end of the tracking. Indeed, cell size is biologically insightful because the fitness landscape of cell size in the absence of death has been shown to reveal sources of stochasticity in volume partitioning and in single-cell growth rate \cite{genthon_analytical_2022}, and to inform on the mechanism of cell size control \cite{thomas_analysis_2018}.

Let us first consider the pre-exposure elongation rate, which is defined by the slope of logarithmic cell size over 1 hour before the INH addition. Among the 274 lineages alive at or dead by $t=72$ h (from drug exposure), 5 lineages have negative pre-exposure elongation rate, which are removed from the analysis. 
The three probability distributions $p_{\rm{for}}^{\star}$, $p_{\rm{for}}$ and $p_{\rm{back}}$ of the pre-exposure single-cell growth rate (for lineages extracted at $t=72$ h) are relatively close to each other, as shown in \cref{fig_tree_landscape_death}B, and equivalently the landscapes $h_t^\star$, $h_t^\dagger$ and $h_t$, shown in \cref{fig_tree_landscape_death}D, are all close to flat landscapes. 
Note that only the dependence of the landscapes on the cell trait is important, and not the offset between the three curves which only comes from the different constants $\Lambda_t$ and $\Gamma_t$ used in the definitions \cref{eq_def_h_death,def_h_dag}. 
These observations are consistent with the analysis in \cite{wakamoto_dynamic_2013} (Fig 2C left in this reference), namely the flatness of the survivor bias $h_t^\dagger$ confirms the independence of the pre-exposure single-cell elongation rate with the viability of the lineage when exposed to drug. 
Further, our framework provides an additional result: the flatness of the fitness landscape $h_t^\star$ indicates the independence between the pre-exposure single-cell elongation rate and the fitness (reproductive success) of the lineage.

Let us now consider the second cell trait, namely the logarithmic cell size at the end of the lineage. We extracted lineages alive at time $t=36$ h and those killed before this time, and measured the logarithmic cell size at the end of the lineages, defined as the extraction time $t=36$ h for surviving lineages and as the time of death for dead lineages. The corresponding probability distributions $p_{\mathrm{for}}$, $p_{\mathrm{for}}^{\star}$ and
$p_{\mathrm{back}}$, and landscapes $h^{\star}_t$, $h_t$ and $h^{\dagger}_t$ are shown in \cref{fig_tree_landscape_death}C and E. 
This analysis reveals two results. All three landscapes are decreasing functions of the final log-size (except for the last point which is not meaningful due to sampling issues), which indicates that cell size is negatively correlated with both fitness and survival. Moreover, the slope of the decrease of $h_t$ is larger than that of the fitness landscape $h^{\star}_t$(or equivalently the difference between $p_{\mathrm{back}}$ and $p_{\mathrm{for}}$ is more important than the one between $p_{\mathrm{back}}$ and $p_{\mathrm{for}}^{\star}$). Therefore, by looking at $h_t$ instead of $h^{\star}_t$ one could incorrectly infer a fitness advantage for smaller cells that is larger than in reality, since the decrease of $h_t$ is mainly due to the survivor bias.

We claim that the decrease of the landscapes for final cell size on \cref{fig_tree_landscape_death}E truly captures fitness and survival effects, while the variations of the same landscapes for the pre-exposure growth rate around their mean values on \cref{fig_tree_landscape_death}D are not meaningful but rather the result of uncorrelated fluctuations. 
To confirm this point, we plot on \cref{fig_preexpgr} in \cref{sec_data} the same landscapes after shuffling the trait values among lineages, thus canceling the correlations between cell trait and number of divisions and between cell trait and survival. 
The resulting landscapes are significantly flatter than the ones in \cref{fig_tree_landscape_death}E for final cell size, while they are compatible with the ones in \cref{fig_tree_landscape_death}D for the pre-exposure growth rate, which confirms our analysis.

\subsection{Inferring the division and death rates}
\label{sec_inference}

\begin{figure}[t]
	\includegraphics[width=\linewidth]{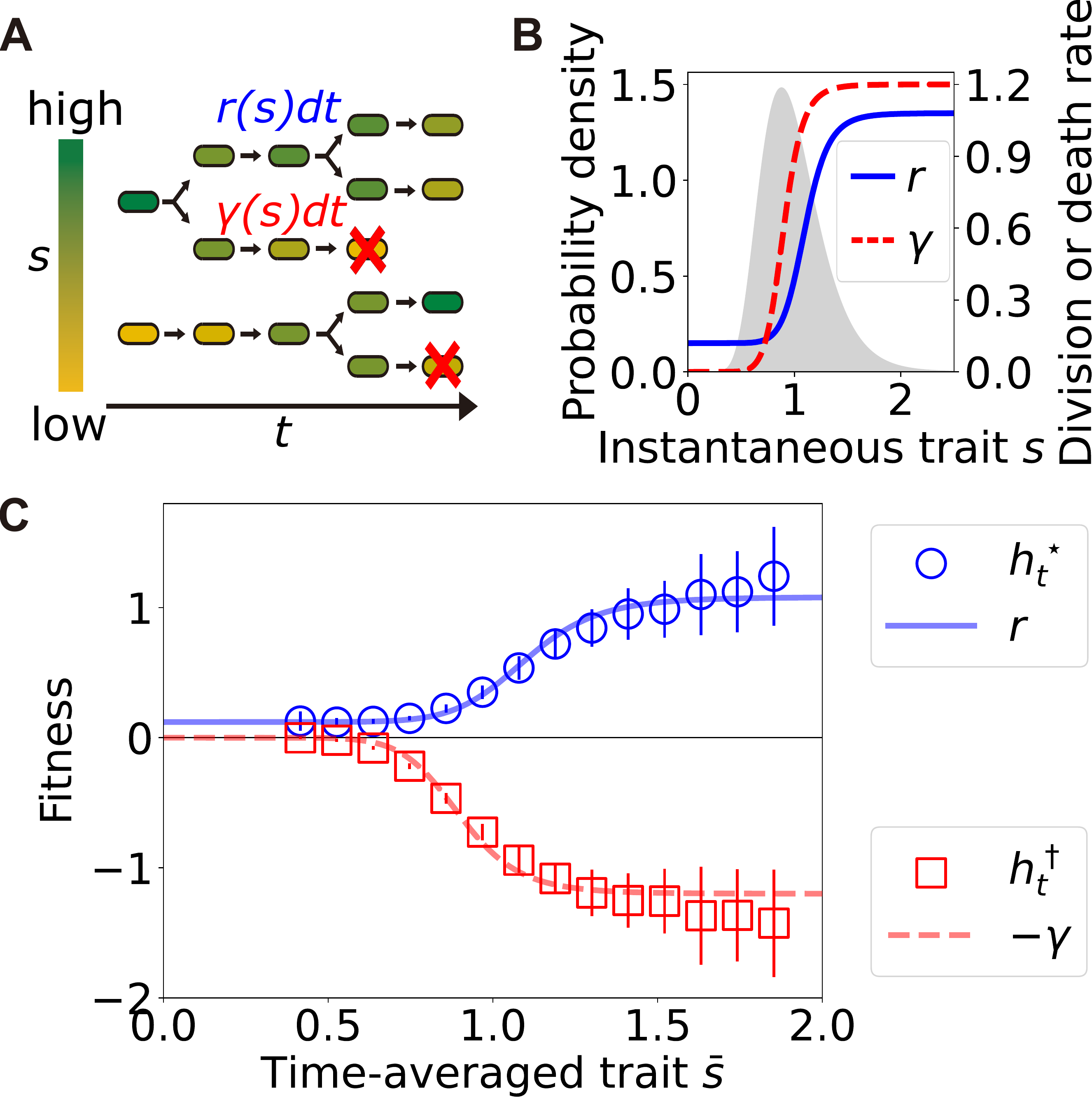}
	\caption{Inference of division and death rates from fitness landscape and survivor bias of time-averaged trait in the numerical simulation.
		(A) A schematic of the population model used in the simulation.
		A cell with the trait $s$ divides with the probability $r(s)\di t$ and dies with the probability $\gamma(s)\di t$ during $\di t$.
		(B) Filled curve: Stationary forward probability distribution of the trait $s$ when $\gamma(s)=0$ (no bias).
		Blue solid and red dashed curves are the division rate $r(s)$ 
		and death rate $\gamma(s)$, given in \cref{sec_correl_simu}.
		(C) Fitness landscape $h_t^*(\bar{s})$ and survivor bias $h_t^\dagger(\bar{s})$ of time-averaged trait $\overline{s}$.
		We choose the parameters so that the population size decreases like what happens in the experimental data we analyzed in \cref{fig_tree_landscape_death}, 
		and we choose $t=4$, which is shorter than the auto-correlation time of $s$ and much shorter than the extinction time scale.
		The points and the error bars represent means and standard deviations over 100 runs, each starting with $N_0 = 1000$ cells.}
	\label{fig_simu}
\end{figure}

In the previous sections, we showed that the fitness landscape and the survivor bias provide a universal framework for the analysis of any incomplete lineage tree regardless of its dynamics. 
If we now introduce a cell trait dynamics, the fitness landscape and the survivor bias can in addition be used for the inference of the division and death rates, under certain conditions detailed below.
This is particularly interesting since in general the estimation of the division rate from lineage data is a hard problem \cite{hoffmann_nonparametric_2016,doumic_nonparametric_2012,doumic_individual_2021}. 

More specifically, let us follow a cell trait $\mathcal{S}$ that fluctuates in time along the lineages, with cells dividing and dying at rates $r(s)$ and $\gamma(s)$, respectively. These rates are defined as follow: a cell with trait value $s$ divides (dies) with probability $r(s) \di t$ ($\gamma(s) \di t$) in a time interval $\di t$.
We show in \cref{sec_leibler} that when (i) the cell trait $s$ is unaltered by divisions, and (ii) either (a) the division rate $r(s)$ is only weakly non-linear or (b) the auto-correlation time of the trait dynamics is large compared to the observation time, then:
\be
h_t^{\star}(\overline{s})=r(\overline{s}) \,,
\ee
where we defined the time average along a lineage $\overline{f}=t^{-1}\int_{0}^{t} \di t' f(t')$.
This result can be used in the context of stochastic gene expression for example, where the concentration of expressed proteins is typically continuous at division and fluctuate very little over multiple generations \cite{lin_disentangling_2021}.

Moreover, when the auto-correlation time of the trait dynamics is large compared to the observation time, then:
\be
h_t^{\dagger}(\overline{s})=-\gamma(\overline{s}) \,.
\ee
These relations show that the division (resp. death) rate can be extracted from the fitness landscape (resp. survivor bias) for the time-averaged trait value $\overline{s}$ along lineages. The two important rates can thus be inferred from phenotypic trajectories only when the above conditions hold.

We put this inferring method to the test by simulating clonal cell proliferation and death processes. The cell trait $\mathcal{S}$ is chosen to follow an Ornstein-Uhlenbeck process, and cells divide and die according to division and death rates given by Hill functions. At division, the mother cell produces $m=2$ daughter cells whose initial states $s$ are the same as that of the mother upon division. A schematic representation of the simulation and the plots of the division and death rates are shown in \cref{fig_simu}A and B. The details of the simulations can be found in \cref{sec_correl_simu}. 

The results of the inference are shown on \cref{fig_simu}C, where the auto-correlation time of the trait dynamics was chosen to be larger than the duration of the experiment. As expected, the inference method is successful in this limit, despite the non linearity of the rates with respect to the value of the trait.

The effects of the non-linearity of the division rate and of the auto-correlation time on the inference of the division rate have been investigated in \cite{nozoe_inferring_2017} in the absence of death, and conclusions are similar with death. 
The impact of the auto-correlation time of the trait dynamics on the inference of the death rate is further investigated in \cref{sec_correl_simu} and \cref{si_fig_simu}, where the same simulation is performed with a shorter auto-correlation time of the trait dynamics.

\subsection{Effect of death on selection}

\subsubsection{Measure of death-induced change in the strength of selection}

We reviewed in \cref{sec_background} that the degree of dissimilarity between the forward and backward distributions for a given cell trait $\mathcal{S}$ is a natural measure of selection acting on this cell trait in the absence of death. This definition was motivated by the fact that the forward sampling is built to suppress the differences in reproductive success among the lineages that are present in the backward sampling. 

To generalize this definition in the presence of death while isolating the strength of selection from survival biases, we compute the strength of selection given in \cref{eq:selstr_nodeath} for surviving lineages only, like what we did for the fitness landscape in \cref{sec_fit_surv}:
\begin{equation}
\label{eq_Pi_diff}
\Pi_{\mathcal{S}} =\langle h_t^{\star} \rangle_{\rm{back}} - \langle h_t^{\star} \rangle_{\rm{for}}^{\star} \,.
\end{equation}

Then, we propose a measure of the death-induced change in the strength of selection:
\begin{equation}
\Delta \Pi_{\mathcal{S}} = \Pi_{\mathcal{S}} - \Pi^{\circ}_{\mathcal{S}} \,, 
\end{equation}
where the superscript $^{\circ}$ in $\Pi^{\circ}_{\mathcal{S}}$ indicates observables in the absence of death.
The sign of $\Delta \Pi_{\mathcal{S}}$ indicates if death increases or decreases the distance between the forward and backward distributions, i.e. the strength of selection.

Interestingly, the interplay between intrinsic selection and survival is explicit when focusing on trajectories $\boldsymbol{\mathcal{S}}$ of traits subject to a division rate $\gamma(s)$. Indeed, we show in \cref{app_Pi_traj} that $\Delta \Pi_{\boldsymbol{\mathcal{S}}}$ can be expressed as:
\be
\label{eq_delta_pi_cov}
\Delta \Pi_{\boldsymbol{\mathcal{S}}}=\frac{{\rm{Cov}_{back}^{\circ}} \lp h_t^{\circ},p_{\rm{surv}} \rp}{\langle p_{\rm{surv}} \rangle_{\rm{back}}^{\circ}}  - \frac{{\rm{Cov}_{for}^{\circ}} \lp h_t^{\circ},p_{\rm{surv}}\rp }{\langle p_{\rm{surv}} \rangle_{\rm{for}}^{\circ}}   \,,
\ee
which depends on the correlations (covariances) between the fitness landscape in the absence of death $h_t^{\circ}(\sbf)$, representing the intrinsic selection effect, and the probability of survival up to time $t$ for a trait trajectory $\sbf$ along a lineage:
\be
\label{p_surv}
 p_{\rm{surv}}(\sbf)=\exp \lbk -\int_{0}^{t} \di t' \gamma(s(t')) \rbk \,.
 \ee

\subsubsection{Illustrative example}

We illustrate the possible outcomes of this change in selection strength by considering a simple birth and death process with two states $a$ and $b$. Cells divide and die with state-dependent division and death rates $r$ and $\gamma$.
To avoid extinction, we suppose that $r_a \geq \gamma_a$ and $r_b \geq\gamma_b$, and we start with a large even number $N_0$ of initial cells, in which phenotypes are equally represented: $N_0(a)=N_0(b)=N_0/2$. Cells cannot switch to the other phenotype, and at division two cells of the same phenotype as the mother are produced. Without loss of generality we suppose that phenotype $a$ divides faster than $b$: $\Delta r = r_a-r_b >0$. 

To treat this case explicitly, let us introduce the cell trait $\mathcal{S}$, taking value $s=1$ for cell state $a$ and $s=0$ for cell state $b$.
The number of cells in the subpopulation $a$ evolves as $n(a,t)=N_0 \exp \lbk t (r_a - \gamma_a) \rbk /2 $ and similarly for the subpopulation $b$,
so the backward probability reads
\begin{equation}
	p_{\rm{back}}(s,t) = \frac{e^{t(r_a-\gamma_a)} \delta(1-s) + e^{t(r_b-\gamma_b)} \delta(s) }{e^{t(r_a-\gamma_a)} + e^{t(r_b-\gamma_b)} } \,.
\end{equation}
Phenotypes are equally represented in the initial distribution, and the survival probability up to time $t$ for phenotype $a$ is given by $\exp\lp -t \gamma_a \rp$ (and similarly for $b$), therefore the forward distribution reads
\begin{equation}
	p^{\star}_{\T{for}}(s,t) =\frac{e^{-t\gamma_a} \delta(1-s)+  e^{-t\gamma_b}\delta(s)}{ e^{-t\gamma_a} + e^{-t\gamma_b}}   \,.
\end{equation}
The fitness landscape $h_t^{\star}(s)$ is obtained by computing the ratio of these two distributions (\cref{eq_def_h_death}):  
\begin{equation}
h_t^{\star}(s)=r_a \delta(1-s) + r_b \delta(s) \,.
\end{equation}
Note that this is a simple case where the inference method proposed in \cref{sec_inference} is applicable. Indeed, the fitness landscape exactly recovers the division rate, because the trait is not fluctuating in time. 

Finally, the strength of selection is computed using \cref{eq_Pi_diff}:
\begin{equation}
	\label{eq_Pi_S_2_state}
	\Pi_{\mathcal{S}}=\frac{r_a e^{t(r_a-\gamma_a)} + r_b e^{t(r_b-\gamma_b)}}{ e^{t(r_a-\gamma_a)} + e^{t(r_b-\gamma_b)}} - \frac{r_a e^{-\gamma_a t} +  r_b e^{-\gamma_b t}}{e^{-\gamma_a t} + e^{-\gamma_b t}} \,.
\end{equation}
In the absence of death ($\gamma_a=\gamma_b=0$), the asymptotic value of the strength of selection is controlled by the difference in reproductive rate: $\underset{t \rightarrow \infty}{\lim} \Pi_{\mathcal{S}}^{\circ}  = \Delta r/2$.
With death, the asymptotic selection strength results from two processes: the competition for the largest population, controlled by the net offspring production rate $r-\gamma$, which sets the backward trait distribution; and the competition for the smallest death rate $\gamma$ which controls the forward trait distribution.
From \cref{eq_Pi_S_2_state}, the possible outcomes for $\Delta \Pi_{\mathcal{S}}$ are:
\be
\label{eq_delta_pi_ex}
\Delta \Pi_{\mathcal{S}} \underset{t \rightarrow \infty}{\rightarrow} \begin{cases}
	\Delta r/2 & \text{if } \gamma_a > \gamma_b \text{ and } r_a-\gamma_a > r_b-\gamma_b \\
	-\Delta r/2 & \text{if } \gamma_a > \gamma_b \text{ and } r_a-\gamma_a < r_b-\gamma_b \\
	0 & \text{if } \gamma_a > \gamma_b \text{ and } r_a-\gamma_a = r_b-\gamma_b \\
	0 & \text{if } \gamma_a = \gamma_b \\
	-\Delta r/2 & \text{if } \gamma_a < \gamma_b  \,.
\end{cases}
\ee

Death has no effect on selection ($\Delta \Pi_{\mathcal{S}}=0$) in two situations. 
First, when death is uniform: $\gamma_a=\gamma_b$, then each covariance in \cref{eq_delta_pi_cov} is null. 
Second, when $r_a-\gamma_a = r_b-\gamma_b$ (which implies that $\gamma_a>\gamma_b$), meaning that death `shifts' the forward and backward distributions equally, so that their distance remains constant. In this last case, the covariances in \cref{eq_delta_pi_cov} are equal. 
Note that the critical birth-death process where the total population is maintained constant, in contrast to the super-critical case where it grows exponentially, is obtained by setting $r_a=\gamma_a$ and $r_b=\gamma_b$. Thus, in the critical case we always have $r_a-\gamma_a = r_b-\gamma_b$, and $\Delta \Pi_{\mathcal{S}}=0$.

There is only one case where selection is increased by death ($\Delta \Pi_{\mathcal{S}}>0$): when cells that divide faster also die faster, while keeping a larger division-death balance (i.e. $\gamma_a > \gamma_b$ and $r_a-\gamma_a > r_b-\gamma_b$). In that way, phenotype $a$ remains over-represented in the population like in the absence of death ($p_{\rm{back}}(s=1,t) \to 1$), but because it dies faster, its forward weight vanishes ($p_{\rm{for}}^{\star}(s=1,t) \to 0$, against $p_{\rm{for}}^{\circ}(s=1,t) =1/2$ at all times in the absence of death).
As a consequence, the distance between the two samplings increases. 

On the other hand, selection can be decreased by death ($\Delta \Pi_{\mathcal{S}}<0$) in two cases. If the balance between division and death favors phenotype $b$ while maintaining a smaller death rate for $b$ (i.e. $r_a-\gamma_a < r_b-\gamma_b$ and $\gamma_a > \gamma_b$), then phenotype $b$ dominates both samplings, which become identical ($p_{\rm{back}}(s=0,t) \to 1$ and $p_{\rm{for}}^{\star}(s=0,t) \to 1$). On the contrary, when phenotype $b$ has an unfavorable division-death balance and also dies faster (i.e. $\gamma_a<\gamma_b$), it remains under-represented in the population, and also becomes under-represented in the forward sampling, so that the forward and backward distributions become identical ($p_{\rm{back}}(s=0,t) \to 0$ and $p_{\rm{for}}^{\star}(s=0,t) \to 0$). In these two cases, the strength of selection in the presence of death is null, and so the distance between the two samplings is decreased by death.
Note that this death-induced decrease in the strength of selection can be equivalently obtained by a reduction of fitness (here fitness is understood as division rate, since there is no phenotype switching \cite{fisher_genetical_2000}). Indeed, $\Delta \Pi_{\mathcal{S}} = -\Delta r/2$ can be achieved without death by instead lowering $r_a$ to $r_b$.

One take-away message from this analysis is that the strength of selection is more than a difference of growth rates: it instead measures how the intrinsic forward statistics are modified when cells are under selection in a population. In this setting, death rates matter too, and the strength of selection can be increased, decreased or unaffected by death.

\section{Non-equilibrium thermodynamics interpretation}
\label{sec_thermo}

Now that we have motivated our approach by showing the importance of distinguishing between fitness and survival in data analysis, and by proposing a practical use of these two notions for inference purposes, in this section we further analyze our framework from the perspective of stochastic thermodynamics. 
In the past decade, the thermodynamic structure of population dynamics, and of biological branching trees in general, has been thoroughly investigated \cite{mustonen_fitness_2010,kobayashi_fluctuation_2015,sughiyama_pathwise_2015,garcia-garcia_linking_2019,genthon_fluctuation_2020,genthon_universal_2021}. These works provided insights in the principles which govern evolution and selection, with a degree of universality similar to that of thermodynamics for physics. 
In line with these works, we explore here the specific impact of death on these universal principles.

\subsection{Fluctuation theorems}

When comparing the probabilities to pick a living lineage with $K$ divisions in the backward and forward samplings, \cref{eq_def_p_back,eq_def_p_for,eq_def_p_star}, we obtain: 
\be
\label{eq_ft_death}
p_{\rm{back}}(K,t) =p^{\star}_{\T{for}} (K,t) \ e^{ K \ln m - t \lp \Lambda_t - \Gamma_t \rp  }\,.
\ee
Note that trait $s$ has been integrated out here, since the exponential bias does not depend on it. 
This relation is a generalization of the fluctuation theorem obtained for branching tree without death \cite{nozoe_inferring_2017,genthon_fluctuation_2020}, but unlike fluctuation theorems in stochastic thermodynamics which also compare two probability distributions \cite{seifert_stochastic_2012}, the forward and backward distributions are not related by time-reversal symmetry. 
Death modifies the relations in two ways, through the term $\Gamma_t$ and in the lowering of the population growth rate $\Lambda_t$ which is no longer necessarily positive. 

Integrating the forward probability in \cref{eq_ft_death} gives a first integral fluctuation theorem:
\be
\label{eq_lam_back}
\langle e^{t \Lambda_t -K \ln m} \rangle_{\T{back}} = 1 - p_{\T{for}} (\sigma=0,t) \,,
\ee
which is analogous to a generalization of Jarzynski's equality for absolutely irreversible processes \cite{murashita_nonequilibrium_2014}, for which time-reversed processes are never observed.
Similarly here, dead lineages have a positive weight in the forward sampling but a null one in the backward sampling. Thus, in this analogy $K$, $\Lambda_t$ and $p_{\T{for}} (\sigma=0,t)$ play the role of the work, the free energy, and the 
total statistical weight of absolutely irreversible transitions, respectively.
A concrete consequence of this approach is presented in \cref{sec_2nd_law}.

Integrating the backward probability in \cref{eq_ft_death} leads to a second integral fluctuation theorem:
\be
\label{eq_lam}
\Lambda_t=\frac{1}{t} \ln \langle m^K \rangle_{\T{for}}^{\star} + \Gamma_t \,,
\ee
which is a generalization of \cref{eq:popfitness_nodeath} in the presence of death. This relation links the population growth rate to the forward statistics of the number of divisions and to the forward probability of survival. This relation clearly shows that an effect of death can be equivalently described by a reduction of fitness: since $\Gamma_t <0$, the population fitness $\Lambda_t$ is reduced; which could also be obtained by a reduction of the number of divisions $K$ along the lineages. This death-induced reduction of population fitness is analyzed using the framework of fluctuation-response relations in the next section.

\subsection{Fluctuation-response relation for the population growth rate}

When comparing experiments with and without death, it is natural to ask whether it is possible to predict the effect of death on population growth. 
In this direction, Yamauchi et al. \cite{yamauchi_unified_2022} recently showed that for the particular case in which cell death occurs upon division only, and for a small death probability, then at first order the decrease in population growth rate due to death depends on the Kullback-Leibler divergence between the backward and forward distributions of the number of divisions in the absence of death. The Kullback-Leibler (KL) divergence is a positive non-symmetrical information-theoretic distance between probability distributions:
\be
\mathcal{D}_{\T{KL}}(p || q) = \int \di x \ p(x) \ln \frac{p(x)}{q(x)} \geq 0 \,.
\ee

In \cref{app_lrr}, we extend this result and show that for a general death rate $\gamma$, the decrease in population growth rate is given by: 
\be
\label{eq_diff_lambda}
\Lambda_t-\Lambda^{\circ}_t = \frac{1}{t} \ln \langle p_{\T{surv}} \rangle_{\T{back}}^{\circ} \,,
\ee
where the survival probability $p_{\T{surv}}(\sbf)$ for a trait trajectory $\sbf$ along a lineage is given by \cref{p_surv}.
In particular, if death occurs only at division with probability $1-2^{-\epsilon}$, then the survival probability along a lineage with $K$ divisions is $p_{\rm{surv}}(K)=2^{-\epsilon K}$, and for small $\epsilon$ we recover the result from \cite{yamauchi_unified_2022}.

If the death rate is small, i.e. scaled by a small parameter $\epsilon$, the survivor probability in \cref{eq_diff_lambda} can be expanded such that:
\begin{equation}
\label{LR3}
\Lambda_t-\lp \Lambda^{\circ}_t - \epsilon \langle \overline{\gamma} \rangle_{\T{back}}^{\circ} \rp =  \frac{t}{2} \rm{Var}_{\T{back}}^{\circ} \lp \overline{\gamma} \rp \epsilon^2 + O(\epsilon^3) \,.
\end{equation}
When there is no variability in the time-averaged death rate, the population growth rate is simply reduced by the uniform time-averaged death rate: $\Lambda_t = \Lambda^{\circ}_t - \epsilon \overline{\gamma}_0 $, with $\overline{\gamma}_0=\langle \overline{\gamma} \rangle_{\T{back}}^{\circ}$.
If now we consider the case where there is variability while keeping the same average $\langle \overline{\gamma} \rangle_{\T{back}}^{\circ}$, then $\Lambda_t-\lp \Lambda^{\circ}_t - \epsilon \langle \overline{\gamma} \rangle_{\T{back}}^{\circ} \rp$ is the difference between the growth rates with and without variability in death rate, for the same average death rate. This difference is proportional to the variance in death rate. \Cref{LR3} can thus be interpreted as a fluctuation-response relation, where the response is understood as the gain in population fitness which is caused by  an increase of the fluctuations in death rate. 

This result can be put in parallel with other fluctuation-response theorems in evolution, like Fisher's fundamental theorem of natural selection, which states in its simplest form that the time-derivative of the population fitness $\di \Lambda_t / \di t$ is equal to the variance of division rates ${\rm{Var}}(r)$ \cite{fisher_genetical_2000}
Another example of relation of this kind is given in \cite{leibler_individual_2010}, where the time-integrated division rate along a lineage is called historical fitness $H_t(\sbf)=\int_{0}^{t} \di t' \ r(s(t'))=t \overline{r}$. In this work, the authors derived the equality between the increase in mean historical fitness $\partial_{\beta} \langle H_t \rangle$, following the amplification of the differences in division rates $r(s) \to \beta r(s)$, and the variance of historical fitness ${\rm{Var}}(H_t)$. 

In contrast to these other fluctuation-response relations, our result shows the importance of fluctuations of death rate in understanding the change in population growth rate.

\subsection{Second-law-like inequalities bounding the population growth rate}
\label{sec_2nd_law}

\begin{figure}[t]
	\includegraphics[width=\linewidth]{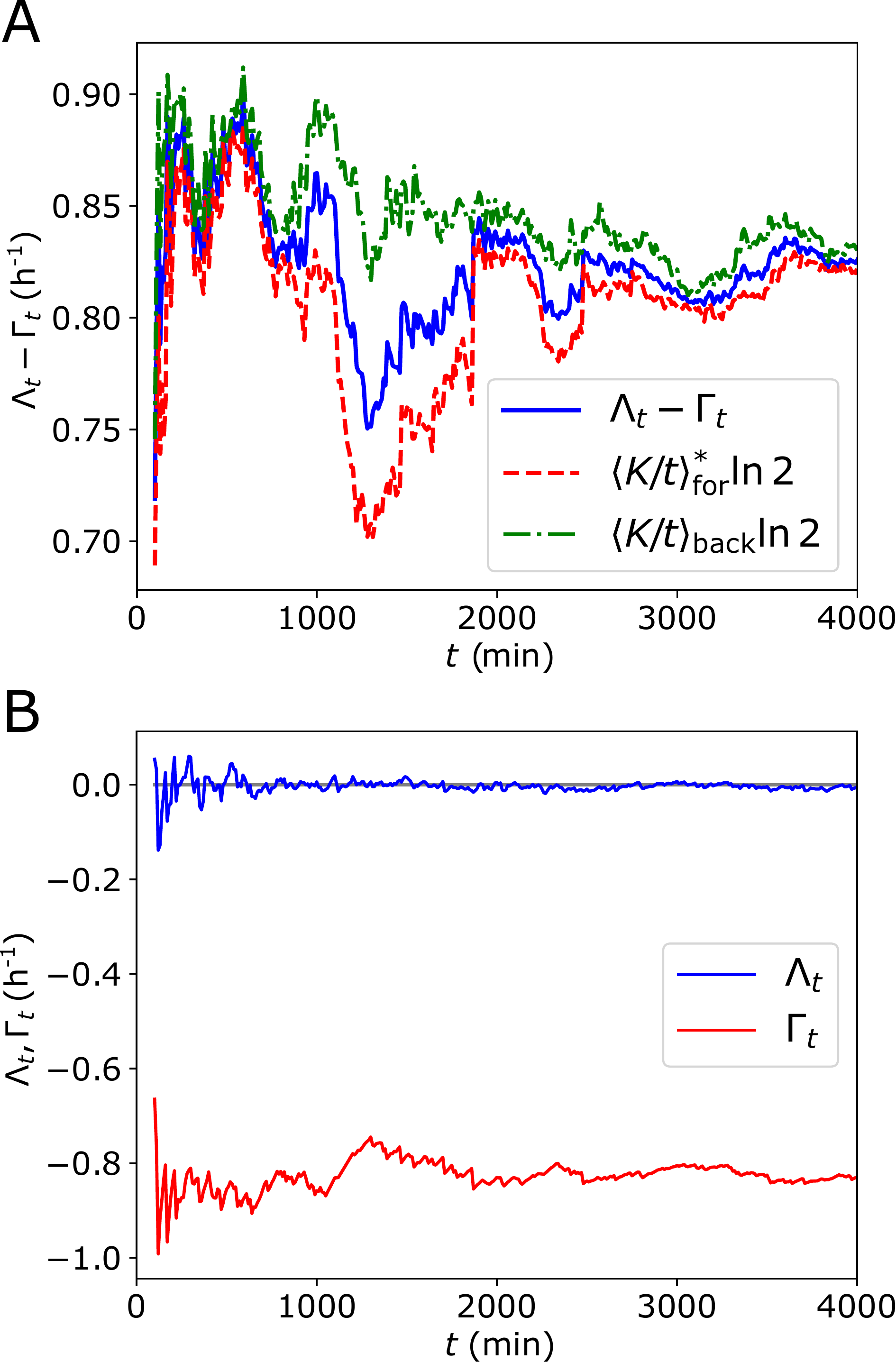}
	\caption{Analysis of data from \cite{hashimoto_noise-driven_2016} for a population maintained constant thanks to dilution. (A) Illustration of \cref{eq_K_ineq}, which shows that the bounds for $\Lambda_t-\Gamma_t$ get tighter as time increases. (B) Separate time evolutions of $\Lambda_t$ and $\Gamma_t$, which show the expected convergence of $\Lambda_t$ toward $0$, due to the fact that the population is maintained constant, and a slower convergence of $\Gamma_t$ toward a steady value, not fully achieved within the reach of the experiment.}
	\label{fig_ineq_death}
\end{figure}%

An important consequence of fluctuation theorems in stochastic thermodynamics is obtained by computing the KL divergence between the original and time-reversed path probabilities, which gives the second-law of thermodynamics. Similarly here, second law-like inequalities can be deduced from the positivity of the following Kullback-Leibler divergences:
\begin{align}
	\mathcal{D}_{\T{KL}}(p_{\rm{back}} || p^{\star}_{\T{for}})
	&= \langle K \rangle_{\T{back}} \ln m - t \lp \Lambda_t - \Gamma_t \rp  \geq 0 \\
	\label{eq_2ndlaw_for}
	\mathcal{D}_{\T{KL}}(p^{\star}_{\T{for}} || p_{\rm{back}}) 
	&= - \langle K \rangle_{\T{for}}^{\star} \ln m + t \lp \Lambda_t - \Gamma_t \rp  \geq 0 \,.
\end{align}
Since the number $K$ of divisions is positive definite, \cref{eq_2ndlaw_for} implies that $\Lambda_t - \Gamma_t$ is a positive quantity.

When combined, the two inequalities give:
\be
\label{eq_K_ineq}
\frac{ \langle K \rangle_{\T{for}}^{\star} \ln m }{t} \leq  \Lambda_t - \Gamma_t  \leq \frac{ \langle K \rangle_{\T{back}} \ln m}{t} \,,
\ee
which provides universal lower and upper bounds for the population growth rate independent of the dynamics.
Let us make two comments. First, \cref{eq_K_ineq} involves two bounds instead of one in thermodynamics, again because the forward and backward samplings are not related by a time-reversal symmetry. Second,
an important result of Ref. \cite{murashita_nonequilibrium_2014} is that the lower bound on the average entropy production is improved due to the irreversible trajectories, going from $0$ to a strictly positive quantity. 
Similarly here, since $\Gamma_t\leq 0$, the right part of \cref{eq_K_ineq} implies $ t \Lambda_t\leq t \lp \Lambda_t - \Gamma_t \rp \leq \langle K \rangle_{\T{back}} \ln m$, giving a tighter bound on the population growth rate. 

We illustrate these inequalities on \cref{fig_ineq_death}A with experimental data from \cite{hashimoto_noise-driven_2016} of {\it Escherichia coli} cells in a microchannel. 
The values of $\langle K \rangle_{\T{for}}^{\star}$, $\langle K \rangle_{\T{back}}$, $\Lambda_t$ and $\Gamma_t$ are computed for a single population, maintained approximately constant with $20\sim40$ cells at any time. We see that the relative discrepancies between the curves are decreasing with time from $t=1300 \ \text{min}$, meaning that the bounds are getting tighter. This is due to the fact that, because of dilution, all cells in the cytometer
are likely to have a common ancestor which is close in the past, thus leading to a small variability in the numbers of divisions amongst lineages \cite{jafarpour_evolutionary_2022}.
We expect each curve to converge to a steady value in the long time limit. 
Fig \ref{fig_ineq_death}B displays the separate evolution of $\Lambda_t$ and $\Gamma_t$: as expected for constant populations, $\Lambda_t$ tends to $0$ rather quickly, while the convergence of $\Gamma_t$ is slower.

When studying regulated populations of cells, \cref{eq_K_ineq} further implies inequalities between the forward and backward mean generation times with the population doubling time, as explored in the next section.

\section{Effect of death on models of cell size control}
\label{sec_csc}

In recent years, the comparison between the forward and backward samplings of lineages has been used to quantify how the behavior of cells is modified when they are under selection. For example, it has been shown that cells elongate faster \cite{thomas_single-cell_2017}, are born bigger \cite{thomas_analysis_2018}, and end up being smaller in a snapshot \cite{genthon_analytical_2022}, 
when selection is present among lineages. Importantly, they also divide faster, because cells that are over-represented in the backward sampling because of their higher reproductive success have shorter inter-division times (also called generation times) \cite{thomas_making_2017,sughiyama_fitness_2019,genthon_fluctuation_2020}. Therefore, selection biases the distribution of cell lifetimes towards small values. 

In the next sections, we introduce a non-interacting age-and-size-structured model of cell populations, and show how this forward-backward bias on generation times is impacted by cell removal due to death or dilution.

\subsection{General model of cell size control with death}

We consider a general model of cell size control, where cells are characterized by their age $a$ and size at birth $x_b$, and divide and die with division and death rates $r(a,x_b)$ and $\gamma(a,x_b)$ respectively. For simplicity, in the following we call $\gamma$ the death rate but keep in mind that it models any kind of cell removal. Upon division, mother-daughter correlations are described by the transition kernel $\Sigma(x_b|x_b',a')$, which is the probability for the daughter cell to be born with size $x_b$ knowing its mother was born with size $x_b'$ and divided at age $a'$. Between divisions, cells grow at a rate $\nu(x)=\di x/\di a$ where $x$ is the cell size. This growth rate accounts for the most common growth laws, for example exponential growth $x(a)=x_b \exp(\nu_0 a)$ when $\nu(x)=\nu_0 x$, and linear growth $x(a)=x_b+ \nu_1 a$ when $\nu(x)=\nu_1$. 
Detailed equations for this model can be found in \cref{app_csc_model}.

In the absence of death, this model accounts for the most common mechanisms of cell size control: the timer, the sizer and the adder, where division is triggered by age, size and increment of volume, respectively \cite{jun_fundamental_2018,cadart_physics_2019}. Indeed, cell size is determined by age, size at birth and single-cell growth rate: $x(a,x_b,\nu)$, and the increment of volume since birth is defined as $\Delta=x-x_b$. Therefore, the timer, sizer and adder models are recovered when respectively setting $r(a,x_b) \equiv \breve{r}(a)$, $r(a,x_b) \equiv \hat{r}(x)$ and $r(a,x_b) \equiv \nu(x)\zeta(\Delta)$, where $\zeta(\Delta)$ is the division rate per unit volume \cite{taheri-araghi_cell-size_2015}. The complete mappings with the sizer and adder are presented in \cref{sec_size_control}.
The model is then extended to account for death by adding a general death rate $\gamma(a,x_b)$.

The division and death rates can be linked to the population growth rate $\Lambda_t$ and to the rate of decrease of the forward probability of survival $\Gamma_t$ defined in \cref{sec_back_for}.
Indeed, we show in \cref{app_Lam_Gam_avg} that when averaging over many population trees, which is equivalent to taking the limit $N_0 \to \infty$ since each initial cell produces an independent tree, the stochastic quantities $\Lambda_t$ and $\Gamma_t$ converge to their expected values:
\begin{align}
	\underset{N_0 \to \infty}{\lim}\Lambda_t &= \overline{\langle (m-1) r -\gamma \rangle_{\rm{back}}} \\
	\label{eq_link_Gam_gam}
	\underset{N_0 \to \infty}{\lim}\Gamma_t &= - \overline{\langle \gamma \rangle_{\rm{for}}^{\star}} \,,
\end{align}
where we used the time-average $\overline{f}=t^{-1}\int_{0}^{t} \di t' f(t')$.

Finally, this model accounts for the different experimental setups. In micro-channels, a uniform dilution rate balancing cell divisions on average is typically assumed \cite{powell_growth_1956,hashimoto_noise-driven_2016,levien_interplay_2020}: $\gamma(a,x_b) \equiv \gamma=(m-1) \langle r \rangle_{\T{back}}$, so that the population is maintained constant ($\Lambda_t=0$). 
In morbidostats, the cell population is instead maintained constant thanks to a control of the death rate \cite{toprak_evolutionary_2012}. 

Note that this model is a particular case of the multitype-age model introduced in \cite{sughiyama_fitness_2019}, where the type is here the newborn size. Therefore, beyond cell size control, the results derived in the next section are equally valid with a general cell type, which could represent a genotypic or phenotypic trait instead of the newborn size.

\subsection{Forward-backward bias for generation times in the presence of death}

In the long time limit, if the population does not go extinct and if it reaches a steady state, the population grows exponentially with a rate $\Lambda= \lim\limits_{t \rightarrow \infty} \Lambda_t \geq 0$, called the Malthus parameter. In this limit and for cells undergoing binary fission ($m=2$), the number of cells doubles after a time $T_d=\ln 2 / \Lambda$, called the population doubling time, and the forward survival probability is reduced by half after a time $T_s=-\ln 2/\Gamma$, with $\Gamma= \lim\limits_{t \rightarrow \infty} \Gamma_t$.
Moreover, when a large-deviation principle for the rescaled number of divisions $K/t$ is observed \cite{levien_large_2020,pigolotti_generalized_2021}, then 
$\lim\limits_{t\rightarrow \infty} \langle K \rangle_{\T{back}}/t = \langle \tau \rangle_{\T{back}}^{-1}$ where $\tau$ is the generation time, and similarly for the forward average. Such principle has indeed been derived for the model of cell size control introduced above \cite{sughiyama_explicit_2018}. Then, in the long time limit, \cref{eq_K_ineq} turns into
\be
\label{eq_tau_ineq}
\langle \tau \rangle_{\T{back}}
\leq \lp T_d^{-1} + T_s^{-1} \rp^{-1} \leq 
\langle \tau \rangle_{\T{for}}^{\star} \,,
\ee
which generalizes the inequalities between population doubling time and mean generation times known for age-controlled populations in absence of death \cite{powell_growth_1956,hashimoto_noise-driven_2016,genthon_fluctuation_2020}. Note that the inequality $\langle \tau \rangle_{\T{back}} \leq \langle \tau \rangle_{\T{for}}^{\star} $ is not modified by death (except for the conditioning on survival in the forward sampling), which was expected since this inequality results from differences in the selection only. On the other hand, the middle term explicitly depends on the forward survival ‘halving time', and $T_{\rm{eff}} = 1/ \lp T_d^{-1} + T_s^{-1} \rp$ represents an effective population doubling time. When there is no death, $T_{\rm{eff}} =  T_d$ is the true population doubling time, and when there is no growth, $T_{\rm{eff}} =  T_s$  is also linked to the population doubling time in microchannel experiments. Indeed, if the population is maintained constant thanks to a dilution rate $\gamma=-\Gamma$ that exactly balances divisions, then $-\Gamma$ is also the rate at which the population would grow in the absence of dilution. 

To go beyond average generation times, we next propose a generalization of the bias between generation time distributions, namely Powell's relation. 
Powell's relation and Euler-Lotka equation are two important results from the literature on age-controlled populations in steady-state. 
In a series of papers, Powell derived the statistical bias between the distributions of generation time when measured in population versus single lineages, first for uncorrelated divisions and then in the presence of Markovian correlations \cite{powell_growth_1956,powell_note_1964,powell_generation_1969}.
Euler-Lotka equation expresses the integral relationship between the population growth rate and the single lineage distribution of generation times \cite{bacaer_short_2011}.
These results have since been generalized in several directions: for models with an age-dependent death rate \cite{lebowitz_theory_1974}, and more recently, for models with non-Markovian correlations \cite{pigolotti_generalized_2021}, for constant populations with a uniform dilution rate \cite{levien_interplay_2020}, and for unbalanced (non-steady-state) growth \cite{jedrak_generalization_2022}.  

We define the steady-state distributions of generation times conditioned on size at birth as 
\be
f(\tau|x_b)= \frac{r(\tau,x_b) p(\tau,x_b)}{\int \di \tau' \ r(\tau',x_b) p(\tau',x_b)} \,,
\ee
where $p(a,x_b)$ is the steady-state joint distribution of age and size at birth, evaluated either in the backward sampling or forward sampling conditioned on survival. This distribution measures the proportion of cells dividing with age $\tau$ and size at birth $x_b$ among all cells dividing with size at birth $x_b$, in either of the samplings. 
In \cref{app_powell}, we derive analytical expressions for $f_{\rm{back}}(\tau | x_b)$ and $f^{\star}_{\rm{for}}(\tau | x_b)$,  and their relationship:
\be
\label{powell_cond_main}
f_{\rm{back}}(\tau | x_b)=  \frac{m f^{\star}_{\rm{for}}(\tau | x_b) e^{-(\Lambda-\Gamma) \tau}}{Z(x_b)}  \,,
\ee
where $Z(x_b)$ is a normalizing factor. Since we showed with \cref{eq_2ndlaw_for} that $\Lambda-\Gamma \geq 0$, then for any newborn size $x_b$, the backward distribution is biased towards small division times, where again $\Lambda-\Gamma$ can be interpreted as an effective population growth rate.

A similar but different relation has been derived previously, which compares instead the backward dynamics with death and the forward dynamics without death \cite{sughiyama_fitness_2019}. We propose in \cref{app_sughi} a complete comparison between these two relations. We believe that \cref{powell_cond_main} has the following advantages over that other formulation. 
First, it compares two distributions evaluated in the same experimental setup, with the same dynamics in the presence of death, which makes sense in light of the definition of fitness landscape proposed in \cref{sec_fit}. 
Second, we obtained an explicit expression for the constant $Z(x_b)$ in terms of the distributions of newborn sizes $x_b$ at birth and at division (see \cref{app_powell}).

Finally, in the absence of mother-daughter correlations, i.e. when the correlation kernel is diagonal $\Sigma(x_b|x_b',a') \equiv \hat{\Sigma}(x_b)$, a simpler version of the relation is obtained: $f_{\rm{back}}(\tau)= m  f^{\star}_{\rm{for}}(\tau) e^{-(\Lambda-\Gamma) \tau}$, where the newborn size $x_b$ does not appear explicitly anymore.
This relation further simplifies for uncorrelated age models without death, in which case the regular Powell’s relation is recovered, namely  $f_{\T{back}}(\tau)=m f_\T{for}(\tau) e^{- \tau \Lambda}$ \cite{powell_growth_1956,hashimoto_noise-driven_2016}.

\subsection{Death can conceal the adder property}
\label{sec_adder}

The adder is a mechanism of cell size control that postulates that cells divide after growing by a certain amount of volume which is independent of their volume at birth. Experimental evidences of this mechanism for bacteria have been provided in two independent simultaneous articles \cite{campos_constant_2014,taheri-araghi_cell-size_2015}. This property has since been observed for many other organisms \cite{willis_sizing_2017,jun_fundamental_2018}, among which yeasts \cite{soifer_single-cell_2016} and archaea \cite{eun_archaeal_2017}. 
It is therefore increasingly seen as the most relevant model of cell size. 

Within the framework of \cite{taheri-araghi_cell-size_2015}, in the absence of death, the adder mechanism is characterized by the independence of the lineage distribution of added volume $\Delta_d$ between birth and division from the newborn volume: $f_{\rm{lin}}(\Delta_d|x_b) \equiv f_{\rm{lin}}(\Delta_d)$. The lineage distribution is defined along independent single lineages, in a mother machine setup \cite{wang_robust_2010}. 
Importantly, in the absence of death and for large samples, the lineage distribution is equivalent to the forward distribution, which is also unbiased by the differences in reproductive success by construction \cite{levien_large_2020,genthon_fluctuation_2020}. Therefore, the adder property also reads $f_{\rm{for}}^{\circ}(\Delta_d|x_b) \equiv f_{\rm{for}}^{\circ}(\Delta_d)$. 
Recently, evolutionary explanations for the emergence of the adder property 
have gained interest: the idea is that the adder model ensures a larger population growth rate compared to the sizer in the presence of cell death \cite{hobson-gutierrez_evolutionary_2023}. 

Because of its ubiquity in cell biology, and its potential importance in evolution, it appears fundamental to investigate how death might impact the adder property. 
Similarly, population effects can impact the adder property, which is why in the Suppl. Mat. of \cite{taheri-araghi_cell-size_2015}, the authors
observed the adder property with single lineages in mother machines, to `avoid[...] known bias effects related to the speed of reproduction \cite{powell_growth_1956}.' In the presence of death, a new bias is introduced which can also blur the adder property for surviving lineages: the survivor bias. 

We show in \cref{app_adder} that the forward distribution of added volume conditioned on newborn size for surviving lineages is biased with respect to the same distribution in the absence of death as:
\begin{equation}
\label{eq_p_cond_Del_main}
f^{\star}_{\rm{for}}(\Delta_d|x_b)= \frac{f^{\circ}_{\rm{for}}(\Delta_d) }{Y(x_b)} e^{- \Gamma \tau(x_b,\Delta_d)  -\int_{0}^{\Delta_d} \di \Delta  \frac{\tilde{\gamma}(x_b+\Delta,\Delta)}{\nu(x_b+\Delta)}} \,,
\end{equation}
where $Y(x_b)$ is a normalizing constant, $\tilde{\gamma}(x,\Delta)=\gamma(a(x,\Delta),x_b(x,\Delta))$ is the death rate expressed using the variables volume and increment and volume, and $ \tau(x_b,\Delta_d)$ is the generation time for a cell born with volume $x_b$ and which divides with volume $x_b+\Delta_d$, which depends on the growth law $\nu(x)$. 
Exponentially-growing cells ($\nu(x)=\nu_0 x$) divide after a time $\tau(x_b,\Delta_d)=\ln[(x_b+\Delta_d)/x_b]/\nu_0$, for example.

\Cref{eq_p_cond_Del_main} can be understood intuitively as follows. The exponential term involving the death rate $\tilde{\gamma}$ is the survival probability for the specific cycle of a cell born with volume $x_b$ and dividing after adding a volume $\Delta_d$. On the other hand, the term $\exp( \Gamma \tau)$ is the forward survival probability introduced in \cref{eq_def_Gamma}, over the duration $\tau(x_b,\Delta_d)$ of that same cycle, which is an average measure of survival (\cref{eq_link_Gam_gam}). 
The bias in \cref{eq_p_cond_Del_main} involves the ratio of these two probabilities, and thus compares the survival probability of a cell to its ensemble average. If survival is correlated with the volume at birth and the added volume, via the death rate $\tilde{\gamma}(x_b+\Delta,\Delta)$, then this bias results in a biased conditional statistics of added volume when considering surviving lineages. 

Since both terms in the exponential explicitly depend on the size at birth, this demonstrates that the distribution $f^{\star}_{\rm{for}}(\Delta_d|x_b) $ is not independent of the newborn volume in general. Therefore, the adder property would be concealed by analyzing surviving lineages only, even for cells obeying the adder principle in the absence of death. Three simple examples where the independence remains true for surviving lineages are discussed in \cref{app_adder}.

Note that in practice, the independence of added volume from newborn volume has often been tested at the level of the first moment only: $\langle \Delta | x_b \rangle_{\T{for}}^{\circ} = \langle \Delta  \rangle_{\T{for}}^{\circ}$, by checking that the slope of $\langle \Delta | x_b \rangle_{\T{for}}^{\circ}$ vs $x_b$ is $0$ (or equivalently that the slope of $\langle x_d | x_b \rangle_{\T{for}}^{\circ}$ vs $x_b$ is $1$, with $x_d$ the size at division) \cite{campos_constant_2014,soifer_single-cell_2016,priestman_mycobacteria_2017,eun_archaeal_2017}. Our result then suggests that $\langle \Delta | x_b \rangle_{\T{for}}^{\star}$ is not independent of $x_b$ in general.

We expect to observe this effect in the new generation of experimental setups which combine population and single lineage experiments in the same controlled environment, allowing for example accurate measurements of the effect of antibiotics at both the population and lineage levels \cite{bakshi_tracking_2021}.

\section{Discussion}

In this paper, we built a general framework to properly account for dead lineages in a statistical analysis of lineage trees. Dead lineages are broadly understood as lineages ending before the end of the experiment, regardless of the cause: biological death, cell removal, dilution, $\dots$
Our framework is based on the one proposed in \cite{nozoe_inferring_2017} in the absence of death, which relies on the forward and backward samplings of lineages.
In the procedure we proposed, dead lineages are sampled only in the forward manner, given that they do not appear in the population at final time. 
The statistical description then necessarily involves a new quantity $\Gamma_t$ to account for the forward weight of surviving lineages, which modifies the relations between forward and backward distributions. 

We showed how to quantify fitness and selection in the presence of death, whatever the cause of death might be. 
We proposed a general measure of the effect of death on selection whose sign indicates if death tends to increase or reduce the difference between the forward and backward statistics. 
This formalism is well adapted to situations in which a trait of interest is strongly correlated with death. For instance the apparition of a resistant phenotype is correlated with the application of antibiotics which kills a large fraction of a bacterial population \cite{lambert_quantifying_2015}. Similarly, in starved colonies of bacteria, the probability of cell death depends on a maintenance cost of metabolism \cite{woronoff_2020,schink_death_2019}, itself related to the growth rate before exposure to antibiotics \cite{Biselli_2020}. Beyond bacteria, in tissues, there is a fine balance of growth and death of cells. When cell death is induced by overcrowding effects, many traits of interest such as cell size are typically correlated with the death rate \cite{eisenhoffer_crowding_2012,marinari_live-cell_2012}. 
It is interesting to note that in several places, death is effectively equivalent to a reduction of fitness, as in the reduction of the strength of selection of \cref{eq_delta_pi_ex}, in the integral relations of \cref{eq_lam,eq_lam_back}, and in the fluctuation-response relation of \cref{LR3}.

When focusing on models of cell size control, new inequalities between mean generation times, population doubling time and survival ‘halving time'; and new generalizations of Powell's relation followed from our approach. 
These generalizations are useful, in particular in experiments in which sampling individuals from a population can be affected by the survivor bias. In fact, even a standard relation used to characterize the adder model may be hidden if a survivor bias is present. 

The present framework generalizes beyond populations of cells, to similar branched trees in other areas of biophysics. In evolutionary biology, lineage trees can be for instance phylogenetic trees, which contain dead lineages due to species extinction \cite{stadler_recovering_2013}.
Naturally, our framework does not solve a central problem in this field: often we do not know
the precise chronology of events from which the forward statistics could be built. As a result, only the backward sampling can be performed, and selection mechanisms and past history of the population need to be inferred together from snapshot data only. Moreover, relations involving a comparison between the two samplings are valid only for surviving lineages, and thus do not take advantage of the large amount of data from dead lineages. These data can be exploited with the tree sampling \cite{thomas_analysis_2018,levien_interplay_2020,nakashima_lineage_2020}, and linking the tree and forward/backward samplings in the presence of a general death rate would be a logical next step of this work.

\section{Acknowledgments}

We acknowledge J. Unterberger, E. Kussell, A. Amir, Y. Sughiyama, TJ. Kobayashi and Y. Wakamoto for stimulating discussions. A. G. and D. L. received support from the grants ANR-11-LABX-0038, ANR-10-IDEX-0001-02.
T. N. received support from JSPS KAKENHI Grant Number JP21K20672, JST ERATO Grant Number JPMJER1902.

\appendix
\onecolumngrid

\section{Fitness landscape and survivor bias recover division and death rates}
\label{sec_leibler}

In \cite{leibler_individual_2010}, the authors introduced the notion of historical fitness, which is equal to the time-integrated division rate along a lineage:
\be
H_t(\sbf)=\int_{0}^{t} \di t' \ r(s(t')) \,,
\ee
where $s$ is a phenotypic trait, possibly of high dimension, and $\sbf=\{s(t')\}_{t' \in [0,t]}$ is a phenotypic trajectory. They considered a simple model of population dynamics with phenotypic fluctuations, where the evolution of the expected number $n(s,t)$ of cells with trait value $s$ at time $t$ follows the equation:
\be
\label{eq_model_leibler}
\partial_t n(s,t) = r(s) n(s,t) + \int \di s' \ T(s,s') n(s',t) \,,
\ee
where $T(s,s')$ is the rate of phenotype switching from $s'$ to $s$. Importantly here, the divisions and phenotype switchings are independent events, and newborn cells have the same phenotype as their mother. 
We call this dynamics the switching-division model.
With this model, the authors argued that the measure of selection proposed by Fisher in \cite{fisher_genetical_2000}, namely the population variance of the division rate (called fitness or individual fitness in this context), was no longer a good measure of selection. 
Instead, they shifted the perspective from individual cells to individual trajectories, and showed that the variance of historical fitness was more adequate to measure selection, in the sense of gauging the importance of selective differences for the evolution of the population. 

Because the historical fitness is model-dependent and because it relies on the division rate which is not often measurable, in \cite{nozoe_inferring_2017} the authors proposed a new notion of fitness: the fitness landscape $h_t(s)$. The latter is independent of the dynamics of the population, and only relies on phenotypic trajectories: it can thus be evaluated in any branching tree. To show the consistency between the fitness landscape and previous notions of fitness, they proved that, when considering a population governed by the simple model \cref{eq_model_leibler}, the fitness landscape recovers the historical fitness: $h_t(\sbf)=H_t(\sbf)/t$. 
When some conditions are met, detailed in the next section, the historical fitness can be well approximated by the division rate for the time-averaged phenotypic trait: $H_t(\sbf)/t \approx r(\overline{s})$. As a consequence, the fitness landscape can be used to infer the division rate by simply sampling phenotypic trajectories. This idea has been proven successful numerically in \cite{nozoe_inferring_2017}. 

In this appendix, we show that the connection between fitness landscape, historical fitness and division rate holds when adding death to the switching-division model. In addition, we show that the survivor bias defined in the main text can be used in a similar way to infer the death rate.

\subsection{Fitness landscape, historical fitness and division rate}

We consider the switching-division model described by \cref{eq_model_leibler} in which we add a phenotype-dependent death rate $\gamma(s)$ and show that the link between historical fitness and fitness landscape hold for surviving lineages. 

We start from the definition of the fitness landscape under the form of \cref{eq_def2_h_death} and for simplicity we consider that the trait $\mathcal{S}$ takes only discrete values $s \in \{s_i\}_{i=1}^n$.
Since the division rate only depends on the current phenotype of the cell, and since divisions do not alter the phenotype, the number of divisions on different portions of the trajectory are independent:
\be
p_{\T{for}}(K_1, ..., K_n,t|\sbf,\sigma=1)= \prod_{i=1}^{n} p_{\T{for}}(K_i,t|\sbf,\sigma=1) \,,
\ee
where $K_i$ is the number of divisions that occured during the time $t_i$ spent in state $s_i$ (even if this duration is discontinuous), called occupation time. The occupation times sum up as $\sum_{i=1}^{n} t_i =t$. 
In each state, the division rate is constant, so that each term in the product is a Poisson distribution:
\be
p_{\T{for}}(K_i,t|\sbf,\sigma=1) = \frac{(r(s_i)t_i)^{K_i}}{K_i!} e^{-r(s_i)t_i} \,.
\ee
Combining these results, we obtain
\begin{align}
	p_{\T{for}}(K,t|\sbf,\sigma=1)&=\sum_{K_1+...+K_n=K} \ p_{\T{for}}(K_1, ..., K_n,t|\sbf,\sigma=1) \\
	& =e^{-\sum_{i=1}^{n}r(s_i)t_i} \sum_{K_1+...+K_n=K} \ \prod_{i=1}^{n} \frac{(r(s_i)t_i)^{K_i}}{K_i!} \\
	\label{eq_heter_pois}
	& = e^{-\sum_{i=1}^{n}r(s_i)t_i} \frac{(\sum_{i=1}^{n}r(s_i)t_i)^K}{K!} \,,
\end{align} 
where we used the multinomial development to obtain the last line.

Finally, plugging \cref{eq_heter_pois} in \cref{eq_def2_h_death} with $m=2$ leads to:
\begin{align}
	h_t^{\star}(\sbf)&= \frac{1}{t} \sum_{i=1}^{n}r(s_i) t_i \\
	&=\frac{1}{t} H_t(\sbf)
\end{align}

Now, let us detail under which conditions the fitness landscape can be used to infer the division rate. If the division rate is linear, then $ \int_{0}^{t} \di t' \ r(s(t'))/t = r(\overline{s})$ exactly, where $\overline{s}=t^{-1} \int_{0}^{t} \di t' \ s(t')$, independently of the trait dynamics. Therefore we have 
\be
h_t^{\star}(\overline{s})=r(\overline{s}) \,.
\ee
The result remains true if the division rate is only weakly non-linear, even with possibly fast fluctuations of the trait, or if the division rate is significantly non-linear, but with slow fluctuations of the trait (i.e. when the auto-correlation time of the trait dynamics is smaller than the observation time), as observed in the main text. 

Note that this inference method works only when tracking a cell trait that is unaffected by divisions, like with the switching-division model \cref{eq_model_leibler}. With models of cell size control introduced in \cref{Main} for example, the cell trait (size, age, ...) is reset at division, and thus the number of divisions is encoded in the trait trajectory $\boldsymbol{s}$, such that $p_{\T{for}}(K,t|\sbf,\sigma=1)=\delta(K-K[\sbf])$. Therefore, $h^{\star}_t(\sbf)=K[\sbf] \ln 2 /t$, which does not inform on the division rate.

\subsection{Survivor bias and death rate}

In this section we show that, under certain assumptions, the survivor bias given in \cref{def_h_dag} can be used to infer the phenotype-dependent death rate $\gamma(s)$. 

First let us derive the explicit bias between $p^{\star}_{\rm{for}} (\overline{s})$ and $p_{\rm{for}}^{\circ}(\overline{s})$, where the superscript $^{\circ}$ indicates quantities from the experiment without death. When comparing two experiments, identical in all points except for the presence of death in one of them only, through a death rate $\gamma(s)$ depending on cell trait $s$, the expected numbers of lineages following the path $\sbf$ are linked by
\be
\label{eq_n_surv_bias}
n(\sbf)=n^{\circ}(\sbf) \exp \lbk - \int_{0}^{t} \di t' \gamma(s(t')) \rbk \,.
\ee
Upon division of \cref{eq_n_surv_bias} by $n_0^{-1}m^{-K[\sbf]}$ (for simplicity we consider that $n_0=n_0^{\circ}$ here), we obtain
\begin{equation}
	p_{\rm{for}}(\sbf,\sigma=1) = p_{\rm{for}}^{\circ}(\sbf) \exp \lbk - \int_{0}^{t} \di t' \gamma(s(t')) \rbk \,.
\end{equation}
Note that here, no assumption is needed as to whether the cell trait is affected or not by divisions. For simplicity, we considered the case where the trajectory $\sbf$ encodes the number of divisions $K[\sbf]$. If this is not the case, the same result can be obtained by comparing the forward joint probabilities of lineages with path $\sbf$ and $K$ divisions, and then summing over $K$ since the exponential term is independent of the number of divisions.

We then condition the probability in the left hand side on survival: $p_{\rm{for}}(\sbf,\sigma=1)=p^{\star}_{\rm{for}}(\sbf) p_{\rm{for}}(\sigma=1,t)$, where by definition $p_{\rm{for}}(\sigma=1,t)=\exp\lp t\Gamma_t\rp$:
\begin{equation}
	\label{eq_p_surv_bias}
	p^{\star}_{\rm{for}}(\sbf) = p_{\rm{for}}^{\circ}(\sbf) \exp \lbk - \int_{0}^{t} \di t' \gamma(s(t')) -t\Gamma_t\rbk \,.
\end{equation}
Now, when the time-integrated death rate can be replaced by the death rate of the time-integrated phenotype along a lineage: $\int_{0}^{t} \di t' \ \gamma(s(t'))/t \approx \gamma(\overline{s})$, we obtain
\be
\label{bias_star_circ}
p^{\star}_{\rm{for}}(\overline{s}) = p_{\rm{for}}^{\circ}(\overline{s}) \exp \lbk -t\gamma(\overline{s})-t\Gamma_t \rbk \,.
\ee
Finally, when the forward distribution $p_{\rm{for}}(\overline{s})$ of time-averaged phenotypes with death (which includes all lineages, even the dead ones) remains close to the same distribution $p_{\rm{for}}^{\circ}(\overline{s})$ without death: $p_{\rm{for}}(\overline{s}) \approx p_{\rm{for}}^{\circ}(\overline{s})$, we plug \cref{bias_star_circ} into \cref{def_h_dag} to obtain:
\be
h_t^{\dagger}(\overline{s})=-\gamma(\overline{s}) \,.
\ee

Like for the division rate, the hypothesis $\int_{0}^{t} \di t' \ \gamma(s(t'))/t \approx \gamma(\overline{s})$ is verified either when the death rate is only weakly non-linear or when the auto-correlation of the trait dynamics is large. 
The assumption $p_{\rm{for}}(\overline{s}) \approx p_{\rm{for}}^{\circ}(\overline{s})$ reflects that death does not induce an important bias on phenotypic averages along trajectories when considering the ensemble of all lineages, both dead and alive. This is true for large auto-correlation of the trait dynamics, independently of the shape of the death rate, as explored in \cref{sec_correl_simu}.

\subsection{Details on simulations and investigation of the effect of auto-correlation time}
\label{sec_correl_simu}

\begin{figure}
	\includegraphics[width=\linewidth]{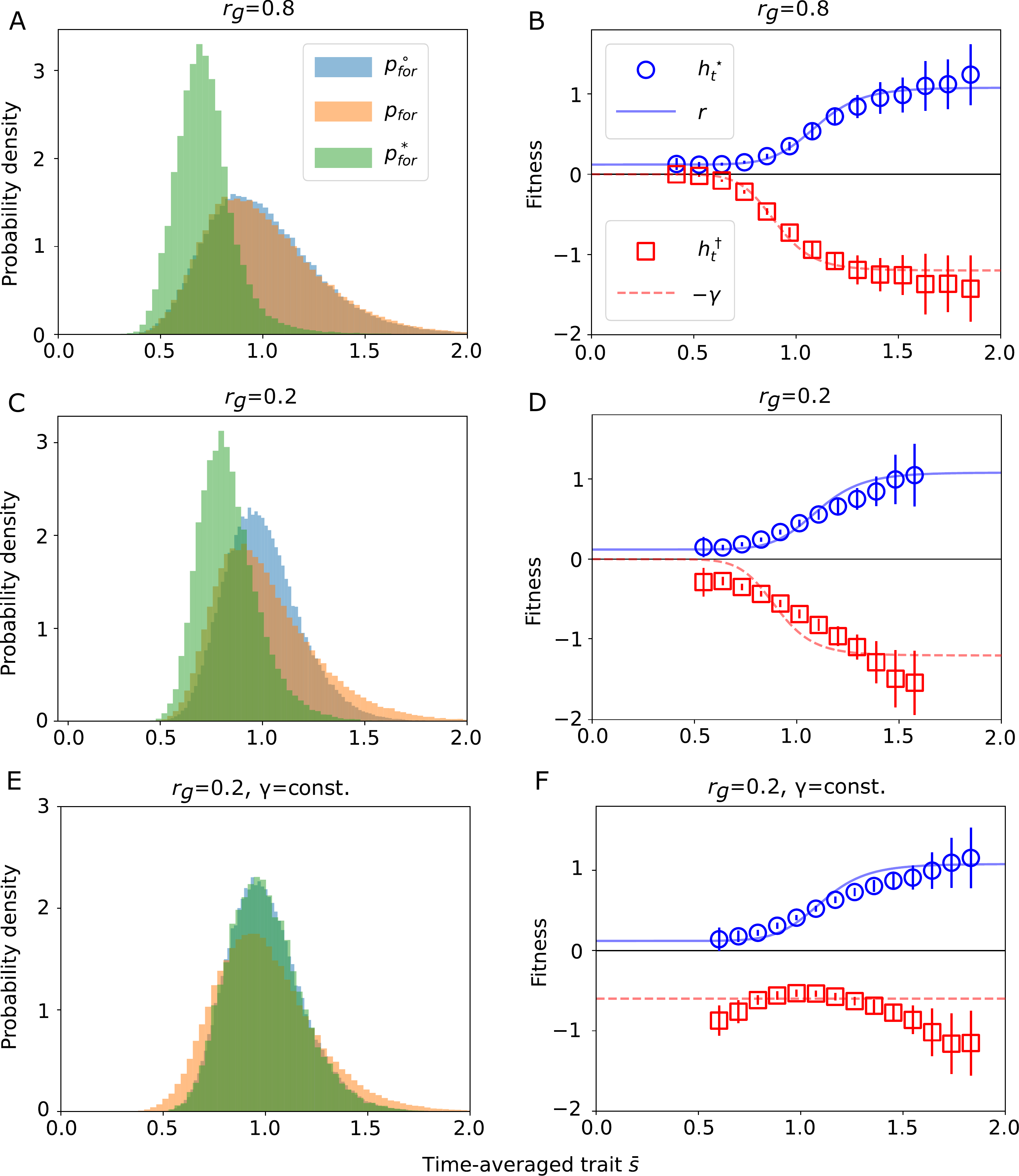}
	\caption{Influence of auto-correlation factor $r_g$ from simulations on the inference method. Left column: forward distributions (each histogram is evaluated from all 100 runs), right column: comparison between the division rate $r(\overline{s})$ (death rate $\gamma(\overline{s})$) and the fitness landscape
		$h_t^\star(\overline{s})$ (survivor bias $h_t^\dagger(\overline{s})$).
		(A, B) Large auto-correlation $r_g=0.8$. 
		(C, D, E, F) Small auto-correlation $r_g=0.2$. 
		(A, B, C, D) Death rate given by $\gamma(s) = 1.2 s^{10} / (s^{10} + 0.9^{10})$. 
		(E, F) Flat death rate $\gamma(s)=0.6$.
		(A, B, C, D, E, F) Division rate $r(s) = 0.12 + 0.96 s^{10} / (s^{10} + 1.1^{10})$ and observation time $t=4$.
	}
	\label{si_fig_simu}
\end{figure}

In this appendix, we give the details of the numerical simulation used in \cref{sec_inference}, and further investigate the role of the time scale of cell trait fluctuations. 

In the simulations used for \cref{fig_simu}, 
we chose an Ornstein-Uhlenbeck process for $\ln s_t$, where $\ln s_{t+\Delta t}$ is randomly sampled from the normal distribution with mean $\mu + e^{-\beta \Delta t} \left( \ln s_t - \mu \right)$ and variance $\sigma^2 \left( 1-e^{-2\beta \Delta t} \right)$, with parameters $\Delta t = 0.05$, $\mu = -0.5 \ln (1.09)$, $\sigma^2 = \ln (1.09)$, and $\beta = -0.6 \ln r_g$. With these parameters, without selection due to $r(s)$ and $\gamma(s)$,  $s_t$ follows the log-normal distribution with mean $1.0$ and standard deviation $0.3$ in the steady state.
A cell with the trait value $s$ divides into $m=2$ cells with the probability of  $r(s)\Delta t$ at each time point, where  $r(s) = 0.12 + 0.96 s^{10} / (s^{10} + 1.1^{10})$.
The initial states of two offspring are the same as that of the mother upon division. A cell with trait value $s$ dies with the probability of $\gamma(s)\Delta t$ where  $\gamma(s) = 1.2 s^{10} / (s^{10} + 0.9^{10})$. 

The parameter $r_g$ indicates a strength of auto-correlation of $s_t$ ($0 < r_g <1 $), and $\tau_{\rm{corr}}=1/\beta$ represents the auto-correlation time of the trait dynamics. 
On \cref{fig_simu}C, we showed the results for final time $t=4$ and $r_g = 0.8$ corresponding to $\tau_{\rm{corr}}=1/\beta=-1/(0.6 \ln r_g)=7.47 >t$. The time-scale of the phenotypic variations is thus larger than the observation time and therefore phenotypes fluctuate slowly. 
This implies that the time-averaged division and death rates can be well approximated by the rates for the time-averaged trajectories, and also that the distribution $p_{\rm{for}}(\overline{s})$ recovers $p_{\rm{for}}^{\circ}(\overline{s})$ with good precision, as illustrated in \cref{si_fig_simu}A. 
The reason for that is that for slowly fluctuating traits, the shorter length of dead lineages does not bias the statistics, since phenotypes do not deviate more significantly for living lineages than for dead ones on the time scale of the observation. 
Note that this does not mean that the distribution for surviving lineages $p_{\rm{for}}^{\star}(\overline{s})$ is not biased from $p_{\rm{for}}(\overline{s})$. Indeed, because the death rate is an increasing function of the trait, cells are more likely to die with large trait values of the trait, and therefore $p_{\rm{for}}^{\star}(\overline{s})$ is biased toward smaller cells as compared to $p_{\rm{for}}(\overline{s})$. 
In \cref{si_fig_simu}A and in the other plots in the following, for simplicity, $p^{\circ}_{\T{for}}(\overline{s})$ is evaluated by running the simulations in the same way but with $\gamma(s)=0$ and $m=1$, given that in the absence of death the forward and lineage distributions are equivalent \cite{levien_large_2020,genthon_fluctuation_2020}. \Cref{si_fig_simu}B is the same as \cref{fig_simu}C from the main text, where the inference is successful.

We now run the simulation with the same observation time $t=4$, but with a lower auto-correlation factor $r_g=0.2$, corresponding to an auto-correlation time $\tau_{\rm{corr}}=1.04 <t$, leading to fast fluctuations of the trait. 

First, in \cref{si_fig_simu}CD, we keep the same death rate as in \cref{si_fig_simu}AB and in the main text. 
Because of fast fluctuations, we easily understand why $\int_{0}^{t} \di t' \ r(s(t'))/t \neq r(\overline{s})$ (and $\int_{0}^{t} \di t' \ \gamma(s(t'))/t \neq \gamma(\overline{s})$), which is why the inference of the division rate in \cref{si_fig_simu}D is slightly less accurate than when $r_g=0.8$. More importantly, we observe that 
the survivor bias fails to recover the death rate correctly. This is explained by the significant difference between $p_{\rm{for}}(\overline{s})$ and $p_{\rm{for}}^{\circ}(\overline{s})$ observed in \cref{si_fig_simu}C, induced by fast fluctuations. Indeed, dead lineages are shorter, and can thus explore values of $\overline{s}$ that are inaccessible for living lineages since phenotypes are likely to auto-average in time. 

Second, for completeness, we show in \cref{si_fig_simu}EF the case of low auto-correlation and flat death rate, to illustrate that the accuracy of the inference of the death rate is truly determined by the trait dynamics, and independent of the shape of the death rate, even if linear, unlike what happens for the division rate. 
In this case, we observe in \cref{si_fig_simu}E that $p_{\rm{for}}(\overline{s})$ and $p_{\rm{for}}^{\circ}(\overline{s})$ are still significantly different despite the flat death rate, for the same reason as in the previous paragraph. This translates in a survivor bias that does not recover the flat death rate in \cref{si_fig_simu}F. Moreover, we observe that $p_{\rm{for}}^{\star}(\overline{s})=p_{\rm{for}}^{\circ}(\overline{s})$, which is expected from \cref{eq_p_surv_bias}.

\section{Shuffled landscapes in the antibiotics experiment}
\label{sec_data}

\begin{figure}
	\includegraphics[width=\textwidth]{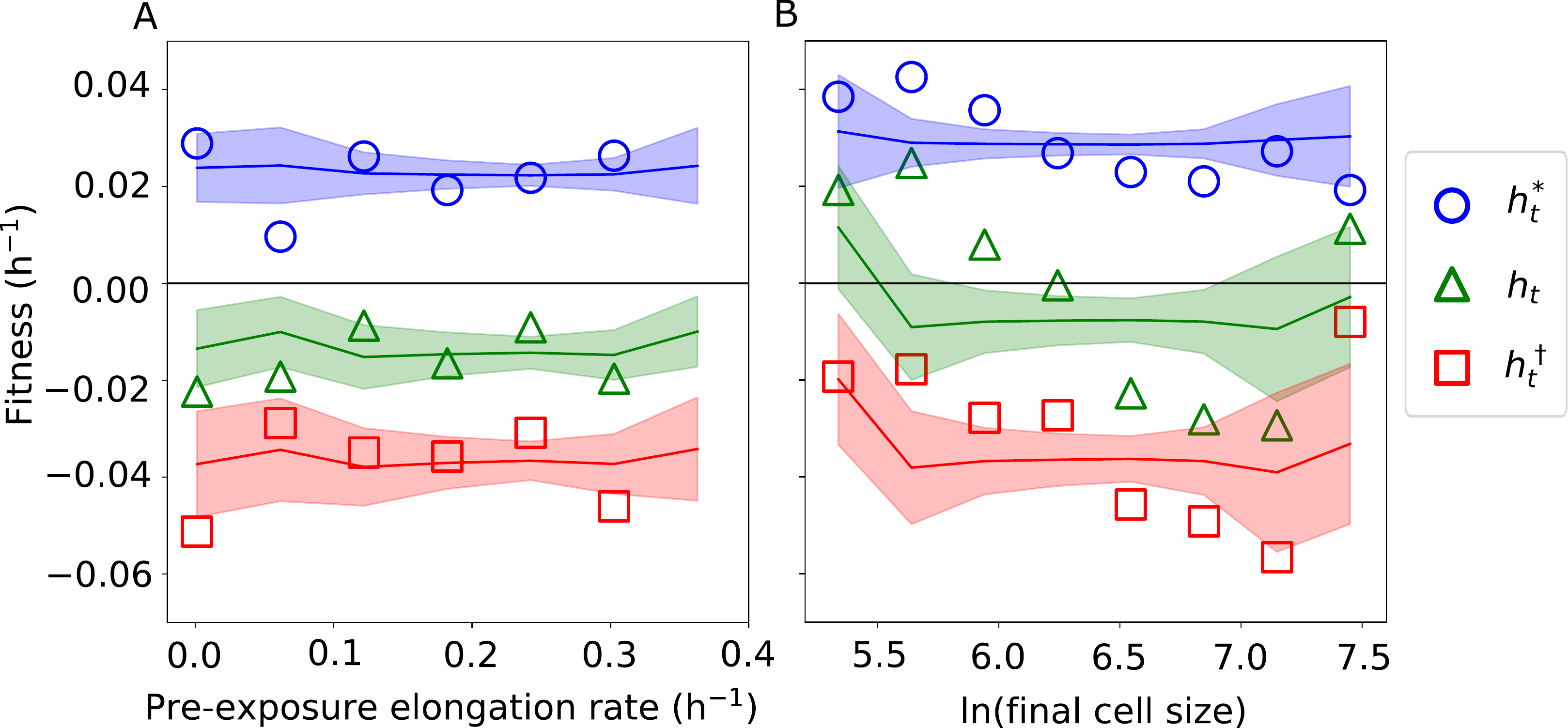}
	\centering
	\caption{Landscapes of (A) pre-exposure elongation rate and (B) final logarithmic cell size, where the curves and the shaded areas show the mean and mean$\pm$standard deviation of each landscape over 10000 random shuffles of the trait values.
	See \cref{sec_data} for the details on the procedure for random shuffling of trait values.}
	\label{fig_preexpgr}
\end{figure}

To evaluate how flat the landscapes $h_t^\star$, $h_t^\dagger$ and $h_t$ are in \cref{fig_tree_landscape_death}, we randomly shuffled the trait values and computed the corresponding new landscapes.
Let $(s_i, K_i, \sigma_i)$ denote the triplet of the trait value (logarithmic final cell size or pre-exposure elongation rate),
the number of divisions, and the fate (0 for dead or 1 for alive) of the $i$-th lineage ($i=1,2,\dots,L$).
Let $\{\rho(i)\}_{i=1,2,\dots,L}$ be a random permutation of $i=1,2,\dots,L$.
A set of triplets with randomly shuffled traits is thus obtained as
$\{ \left(s_{\rho(i)}, K_i, \sigma_i\right)\}_{i=1,2,\dots,L}$.
We shuffled the trait values 10000 times, and for each permutation we computed the landscapes. Then, we calculated the mean and the standard deviation of each landscape, shown as the curves and the shaded areas in \cref{fig_preexpgr}. 
As expected, this procedure cuts the correlations between the trait values and the other values, namely the number of divisions and the survival, leading to flat curves.
By comparing the actual landscapes with the shuffled landscapes, we confirm our analyses. 
Indeed, in \cref{fig_preexpgr}A, the dots lie in the shaded area, showing the independence of the pre-exposure elongation rate with both reproductive success and survival.
On the other hand, in \cref{fig_preexpgr}B for the final logarithmic cell size, the dots are significantly outside of the shaded areas, meaning that the fluctuations of these landscapes really capture fitness and survival effects, rather than random fluctuations.

\section{Effect of death on selection}
\label{app_Pi_traj}

In this appendix, we prove the covariance formula for the strength of selection on trajectories, \cref{eq_delta_pi_cov}. We start from the definition of the strength of selection given in \cref{eq_Pi_diff}: 
\begin{equation}
	\Pi_{\boldsymbol{\mathcal{S}}}  =\frac{1}{t} \int \mathcal{D} \sbf \ \lp p_{\T{back}}(\sbf,t) - p^{\star}_{\T{for}}(\sbf,t) \rp \ln \lp \frac{p_{\T{back}}(\sbf,t)}{p^{\star}_{\T{for}}(\sbf,t)} \rp \,.
\end{equation}
and express $p^{\star}_{\rm{for}}$ and $p_{\rm{back}}$ inside the logarithm using their counterparts in the absence of death. 
To do so, we integrate \cref{eq_n_surv_bias} over trajectories $\sbf$, and we identify the exponential bias as the survival probability \cref{p_surv}, to obtain
\begin{equation}
	\label{eq_surv_bias_N}
	n_t= n_t^{\circ} \ \langle p_{\rm{surv}} \rangle_{\T{back}}^{\circ}  \,.
\end{equation}
Then, we divide \cref{eq_n_surv_bias} by $n_t$ and use \cref{eq_surv_bias_N} to obtain:
\be
\label{eq_surv_bias_back}
p_{\rm{back}}(\sbf)=p_{\rm{back}}^{\circ}(\sbf) \ \frac{p_{\rm{surv}}(\sbf)}{\langle p_{\rm{surv}} \rangle_{\T{back}}^{\circ} }  \,.
\ee
Similarly, the transformation for the forward distribution reads
\be
p^{\star}_{\rm{for}}(\sbf) =p_{\rm{for}}^{\circ}(\sbf) \ \frac{p_{\rm{surv}}(\sbf)}{ \langle p_{\rm{surv}}(\sbf) \rangle_{\rm{for}}^{\circ}} \,,
\ee
which is a rewriting of \cref{eq_p_surv_bias}.

Doing so, the survival probabilities cancel and we obtain:
\begin{align}
	\Pi_{\boldsymbol{\mathcal{S}}} & =\frac{1}{t} \int \mathcal{D} \sbf \ \lp p_{\T{back}}(\sbf,t) - p^{\star}_{\T{for}}(\sbf,t) \rp \ln \lp \frac{p_{\T{back}}^{\circ}(\sbf,t)}{p^{\circ}_{\T{for}}(\sbf,t)} \rp \\
	& = \int \mathcal{D} \sbf \ h_t^{\circ}(\sbf) \lp p_{\T{back}}(\sbf,t) - p^{\star}_{\T{for}}(\sbf,t) \rp  \,.
\end{align}
Then, we subtract $\Pi^{\circ}_{\boldsymbol{\mathcal{S}}} =\int \mathcal{D} \sbf \ h_t^{\circ}(\sbf) \lp p^{\circ}_{\T{back}}(\sbf,t) - p^{\circ}_{\T{for}}(\sbf,t) \rp $ from the relation above:
\begin{align}
	\Delta \Pi_{\boldsymbol{\mathcal{S}}}  = & \int \mathcal{D} \sbf \ h_t^{\circ}(\sbf) \lbk \lp p_{\T{back}}(\sbf,t) - p_{\T{back}}^{\circ}(\sbf,t) \rp  - \lp p^{\star}_{\T{for}}(\sbf,t) - p^{\circ}_{\T{for}}(\sbf,t) \rp \rbk  \\
	= & \frac{1}{\langle p_{\rm{surv}}(\sbf) \rangle_{\T{back}}^{\circ}} \int \mathcal{D} \sbf \ h_t^{\circ}(\sbf) p^{\circ}_{\T{back}}(\sbf,t) \lbk p_{\rm{surv}}(\sbf) - \langle p_{\rm{surv}}(\sbf) \rangle_{\T{back}}^{\circ} \rbk \nonumber \\
	& - \frac{1}{\langle p_{\rm{surv}}(\sbf) \rangle_{\T{for}}^{\circ}} \int \mathcal{D} \sbf \ h_t^{\circ}(\sbf) p^{\circ}_{\T{for}}(\sbf,t) \lbk p_{\rm{surv}}(\sbf) - \langle p_{\rm{surv}}(\sbf) \rangle_{\T{for}}^{\circ} \rbk \\ 
	= & \frac{{\rm{Cov}_{back}^{\circ}} \lp h_t^{\circ},p_{\rm{surv}} \rp}{\langle p_{\rm{surv}} \rangle_{\rm{back}}^{\circ}}  - \frac{{\rm{Cov}_{for}^{\circ}} \lp h_t^{\circ},p_{\rm{surv}}\rp }{\langle p_{\rm{surv}} \rangle_{\rm{for}}^{\circ}}  \,.
\end{align}

\section{Death-induced decrease in population growth rate}
\label{app_lrr}

In this appendix, we derive the general fluctuation-response relation \cref{eq_diff_lambda} and we give an altenative form in terms of a Kullbak-Leibler divergence.  

First, taking the logarithm of \cref{eq_surv_bias_N} leads to
\be
\label{eq_diff_lambda_app}
\Lambda_t-\Lambda^{\circ}_t = \frac{1}{t} \ln \langle p_{\T{surv}} \rangle_{\T{back}}^{\circ} \,.
\ee
In particular, if death occurs only at division with probability $1-2^{-\epsilon}$, then the survival probability along a lineage with $K$ divisions is $p_{\rm{surv}}(K)=2^{-\epsilon K}$. Then for small $\epsilon$, at first order $\ln \langle p_{\T{surv}} \rangle_{\T{back}}^{\circ}=-\epsilon \langle K \rangle_{\T{back}}^{\circ} \ln 2 =-\epsilon \lbk t\Lambda_t^{\circ} + \mathcal{D}_{\T{KL}}(p^{\circ}_{\rm{back}}(K) || p^{\circ}_{\T{for}}(K)) \rbk $, and thus we recover the result from \cite{yamauchi_unified_2022}.

The growth rate difference in \cref{eq_diff_lambda_app} can be expressed as a distance between the statistics with and without death.
To do so, we compute the KL divergence between the backward distributions in the presence and absence of death using \cref{eq_surv_bias_back}:
\be
\mathcal{D}_{\rm{KL}}(p_{\T{back}}^{\circ}(\sbf)||p_{\rm{back}}(\sbf)) = \ln \langle p_{\T{surv}} \rangle_{\T{back}}^{\circ} - \langle \ln  p_{\T{surv}} \rangle_{\T{back}}^{\circ} \,,
\ee
where $\ln  p_{\T{surv}} (\sbf)= - t\overline{\gamma}$ by definition.
The death-induced decrease in population growth rate is therefore given by:
\be
\label{eq_LRR_DKL}
\Lambda_t-\Lambda^{\circ}_t = - \langle \overline{\gamma} \rangle_{\T{back}}^{\circ} + \frac{1}{t} \mathcal{D}_{\rm{KL}}(p_{\T{back}}^{\circ}(\sbf)||p_{\rm{back}}(\sbf)) \,,
\ee
where the first term only depends on the dynamics in the absence of death and the second term is an information-theoretic distance between the statistics with and without death. This relation is similar to recently derived fluctuation-response inequalities valid arbitrarily far from equilibrium which also involve a distance between the statistics before and after a perturbation \cite{dechant_fluctuationresponse_2020,genthon_universal_2021}.

By comparing \cref{eq_LRR_DKL} with \cref{LR3} valid for a small death rate $\epsilon \gamma$, the KL divergence is also linked to the variance in time-integrated death rate:
\begin{equation}
\label{eq_DKL_eps}
\mathcal{D}_{\T{KL}}(p_{\T{back}}^{\circ}(\sbf)||p_{\rm{back}}(\sbf)) = \frac{t^2}{2} \rm{Var}_{\T{back}}^{\circ} \lp \overline{\gamma} \rp \epsilon^2 + O(\epsilon^3) \,.
\end{equation}
Therefore, this divergence is further interpreted as the gain in population fitness due to the variability in time-averaged death rate discussed in the main text.

\section{Details on the models of cell size control}
\label{Main}

\subsection{General model of cell size control with death}
\label{app_csc_model}

In this appendix, we give the equations governing the model of cell size control discussed in \cref{sec_csc}. 
The evolution of the expected number $n(a,x_b,K,t)$ of cells of age $a$, newborn size $x_b$ and $K$ divisions at time $t$ is given by the following Population-Balance Equation (PBE):
\begin{align}
	\label{eq_pop_tim_corr}		
	\partial_t n(a,x_b,K,t) &= - \partial_a  n(a,x_b,K,t) -  \lbk r(a,x_b)  +\gamma(a,x_b) \rbk n(a,x_b,K,t) \\
	\label{eq_pop_tim_BC_corr}
	n(a=0,x_b,K,t) &= m \int \di a' \di x_b' r(a',x_b') \Sigma(x_b|x_b',a') n(a',x_b',K-1,t) \,.
\end{align}
where the second line is the boundary condition for newborn cells, which accounts for divisions of cells with $K-1$ divisions that divide a $K$th time at time $t$ to give birth to cells of size $x_b$.

The PBE and the boundary condition, \cref{eq_pop_tim_corr,eq_pop_tim_BC_corr}, can be recast at the level of the expected backward probability distribution by summing over $K$ and using $p_{\rm{back}}(a,x_b,t)=n(a,x_b,t)/n(t)$ where $n(t)= \int \di a \di x_b \ n(a,x_b,t)$ is the total expected number of cells at time $t$:
\begin{align}
	\label{eq_PBE_death_back}
	\partial_t p_{\rm{back}}(a,x_b,t)= & -\partial_a \lbk p_{\rm{back}}(a,x_b,t) \rbk -\lbk r(a,x_b) +\gamma(a,x_b) + \Lambda_p(t)\rbk p_{\rm{back}}(a,x_b,t) \\
	\label{eq_PBE_death_back_BC}
	p_{\rm{back}}(a=0,x_b,t) =& m \int \di a' \di x_b' \ \Sigma(x_b|x_b',a') r(a',x_b') p_{\rm{back}}(a',x_b',t) \,,
\end{align}
where we defined 
\be
\Lambda_p(t)=\frac{1}{n(t)}\frac{\di n}{\di t} \,,
\ee
the instantaneous population growth rate. 
Direct integration of \cref{eq_PBE_death_back} over $a$ and $x_b$ shows that, in the presence of death, the instantaneous population growth rate is the net difference between the backward-averaged division and death rates:
\be
\label{eq_lam_dyn_death}
\Lambda_p(t) = \int \di a \di x_b \ \lbk (m-1)r(a,x_b) -\gamma(a,x_b) \rbk p_{\rm{back}}(a,x_b,t) \,.
\ee

To recast the PBE for the expected forward probability, we use $p_{\rm{for}}(a,x_b,K,\sigma=1,t)=n(a,x_b,K,t)m^{-K}n(0)^{-1}$ and then sum over $K$, which leads to
\begin{align}
	\label{eq_PBE_death_for}
	\partial_t p_{\rm{for}}(a,x_b,\sigma=1,t)= & -\partial_a \lbk p_{\rm{for}}(a,x_b,\sigma=1,t) \rbk -\lbk r(a,x_b) +\gamma(a,x_b) \rbk p_{\rm{for}}(a,x_b,\sigma=1,t)  \\
	p_{\rm{for}}(a=0,x_b,\sigma=1,t) =&  \int \di a' \di x_b' \ \Sigma(x_b|x_b',a') r(a',x_b') p_{\rm{for}}(a',x_b',\sigma=1,t) \,.
\end{align}
Integrating \cref{eq_PBE_death_for} over $a$ and $x_b$ gives
\be
\partial_t p_{\rm{for}}(\sigma=1,t)= - p_{\rm{for}}(\sigma=1,t) \int \di a \di x_b \ \gamma(a,x_b) p_{\rm{for}}(a,x_b,t|\sigma=1) \,.
\ee
We then define the instantaneous decrease rate of the forward probability of survival:
\be
\label{eq_gam_dyn_death}
\Gamma_p(t) = \frac{1}{p_{\rm{for}}(\sigma=1,t)}\frac{\di p_{\rm{for}}(\sigma=1)}{\di t} = - \int \di a \di x_b \ \gamma(a,x_b) p_{\rm{for}}(a,x_b,t|\sigma=1) \,.
\ee

Finally, one derives the PBE for the expected forward distribution conditioned on survival: $p^{\star}_{\rm{for}}(a,x_b,t)=p_{\rm{for}}(a,x_b,t|\sigma=1)=p_{\rm{for}}(a,x_b,\sigma=1,t)/p_{\rm{for}}(\sigma=1,t)$:
\begin{align}
	\label{eq_for_tim_corr}		
	\partial_t p^{\star}_{\rm{for}}(a,x_b,t) &= - \partial_a  p^{\star}_{\rm{for}}(a,x_b,t) - \lbk r(a,x_b) + \gamma(a,x_b) + \Gamma_p(t) \rbk p^{\star}_{\rm{for}}(a,x_b,t) \\
	\label{eq_for_tim_BC_corr_death}
	p^{\star}_{\rm{for}}(a=0,x_b,t) &= \int \di a' \di x_b' r(a',x_b') \Sigma(x_b|x_b',a') p^{\star}_{\rm{for}}(a',x_b',t) \,,
\end{align}

Note that \cref{eq_for_tim_corr,eq_for_tim_BC_corr_death} can be obtained from the PBE and boundary condition for the backward distribution \cref{eq_PBE_death_back,eq_PBE_death_back_BC} by setting $m=1$ in the boundary term and replacing $\Lambda_p(t)$ by $\Gamma_p(t)$.

\subsection{Link with the sizer and the adder models}
\label{sec_size_control}

The size $x$ of a cell is determined by its size at birth $x_b$, its age $a$ and the single cell growth rate:
\be
\frac{\di x }{\di a} = \nu(x) \,,
\ee
therefore the model presented in \cref{app_csc_model} accounts for the sizer model, where division is triggered by cell size, and for the adder model, where it is triggered by the increment of volume since birth $\Delta=x-x_b$. Let us show explicitly in this appendix how the sizer and adder PBEs are obtained.

First, we re-parametrize the PBE and the boundary condition for the number $\hat{n}(x,x_b,t)$ of cells of size $x$ at time $t$ born with size $x_b$,  with the following change of variables:
\be
n(a,x_b,t) = \hat{n}(x(a,x_b),x_b,t) \nu(x(a,x_b)) \,,
\ee
so that they read
\begin{align}	
	\label{eq_x_xb}
	\partial_t \hat{n}(x,x_b,t) &= - \partial_x  \lbk \nu(x) \hat{n}(x,x_b,t)  \rbk -  \lbk \hat{r}(x,x_b) + \hat{\gamma}(x,x_b) \rbk  \hat{n}(x,x_b,t) \\
	\label{eq_BC_x_xb}
	\nu(x_b) \hat{n}(x_b,x_b,t) &= m \int \di x' \di x_b' \hat{r}(x',x_b') \hat{\Sigma}(x_b|x',x_b') \hat{n}(x',x_b',t) \,,
\end{align}
where we defined the following rates:
\begin{align}
	\hat{r}(x(a,x_b),x_b) &=r(a,x_b) \\
	\hat{\gamma}(x(a,x_b),x_b) &= \gamma(a,x_b) \\
	\hat{\Sigma}(x_b|x'(a',x_b'),x_b') &= \Sigma(x_b|x_b',a') \,.
\end{align}

In the classical sizer model, division is triggered by cell size only. To generalize this model to a dynamics with death, we further assume that death is also triggered by cell size only. We thus make the following assumptions:
\begin{align}
	\hat{r}(x,x_b) &\equiv \hat{r}(x) \\
	\hat{\gamma}(x,x_b) &\equiv \hat{\gamma}(x) \\
	\hat{\Sigma}(x|x',x_b') &\equiv \hat{\Sigma}(x|x') \,.
\end{align}
We integrate \cref{eq_x_xb} over $x_b$ from $0$ to $x$, where the integration of the derivative term yields a boundary term:
\begin{align}
	\int_{0}^{x} \di x_b \ \partial_x  \lbk \nu(x) \hat{n}(x,x_b,t)  \rbk = \partial_x  \lbk \nu(x) \hat{n}(x,t) \rbk - \nu(x) \hat{n}(x,x,t) \,,
\end{align}
which is expressed by plugging \cref{eq_BC_x_xb}. Finally, the PBE for the expected number $\hat{n}(x,t) = \int_{0}^{x} \di x_b \ \hat{n}(x,x_b,t)$ of cells with size $x$ at time $t$ reads:
\be
\partial_t \hat{n}(x,t) = - \partial_x  \lbk \nu(x) \hat{n}(x,t)  \rbk -  \lbk \hat{r}(x) + \hat{\gamma}(x) \rbk \hat{n}(x,t) + m \int \di x' \hat{r}(x') \hat{\Sigma}(x|x') \hat{n}(x',t) \,,
\ee
which is the sizer equation with a death term.

For the adder, let us make a change of variable from $(x,x_b)$ to $(x,\Delta)$, where $\Delta =x-x_b$
is the added volume since birth. The re-parametrization for the expected number $\tilde{n}(x,\Delta,t)$ of cells with size $x$ and added volume $\Delta$ at time $t$ is then given by
\be
\tilde{n}(x,\Delta(x,x_b),t) = \hat{n}(x,x_b,t) \,,
\ee
and we define the rates in the new system of coordinates:
\begin{align}
	\tilde{r}(x,\Delta(x,x_b)) &=  \hat{r}(x,x_b)  \\
	\tilde{\gamma}(x,\Delta(x,x_b)) &= \hat{\gamma}(x,x_b)  \\
	\tilde{\Sigma}(x|x',\Delta'(x',x_b')) &= \hat{\Sigma}(x|x') \,.
\end{align}
In the adder model without death, the separation of variables for the division rate: 
\begin{equation}
\label{eq_r_adder}
\tilde{r}(x,\Delta)=\nu(x) \zeta(\Delta) 
\end{equation}
where $\zeta$ is the division rate per unit volume, is further assumed to account for the independence between size at birth and added volume between birth and division. Different choices for the death rate $\tilde{\gamma}(x,\Delta)$ are discussed in \cref{app_adder}.

\subsection{Population growth and decay rates as average division and death rates}
\label{app_Lam_Gam_avg}

The rates $\Lambda_t$ and $\Gamma_t$ introduced in the main text are stochastic quantities, defined for any single realisation of the stochastic branching process, and therefore they cannot be expressed explicitly in general. However, when averaging over many population trees, they can be linked to the division and death rates. 

Indeed, let us first compute the time-averaged version of the instantaneous rates $\Lambda_p(t)$ and $\Gamma_p(t)$ introduced in the previous section:
\begin{align}
	\Lambda^{\rm{PBE}}_t &= \overline{\Lambda_p} = \frac{1}{t} \ln \lbk \frac{n(t)}{n_0} \rbk \\
	\label{eq_p_for_gam}
	\Gamma^{\rm{PBE}}_t &= \overline{\Gamma_p}  = \frac{1}{t} \ln p_{\rm{for}}(\sigma=1,t) \,.
\end{align}
where we defined the time-average $\overline{f}=t^{-1}\int_{0}^{t} \di t' f(t')$.
We recall that $n(t)$ and $p_{\rm{for}}(\sigma=1,t)$ are the \textit{expected} number of cells and forward probability of survival. 

On the other hand, a given population starting with $N_0$ cells consists of $N_0$ independent trees, so that $N(t)/N_0$ is the empirical average of the number of cells produced after a time $t$ by a single initial cell:
\be
\frac{N(t)}{N_0}=\frac{1}{N_0} \sum_{i=1}^{N_0} N^{i}(t) \,,
\ee
where $N^{i}(t)$ is the number of cells at time $t$ coming from ancestor cell $i$. Thus, in the limit $N_0 \to \infty$, the empirical average $N(t)/N_0$ converges to the true average $n(t)/n_0$, and $\Lambda_t$ and $\Gamma_t$ converge to $\Lambda^{\rm{PBE}}_t$ and $\Gamma^{\rm{PBE}}_t$. Combining this result with \cref{eq_lam_dyn_death,eq_gam_dyn_death}, we obtain:
\begin{align}
	\underset{N_0 \to \infty}{\lim}\Lambda_t &= \Lambda^{\rm{PBE}}_t =\overline{\langle (m-1) r -\gamma \rangle_{\rm{back}}} \\
	\underset{N_0 \to \infty}{\lim}\Gamma_t &= \Gamma^{\rm{PBE}}_t =- \overline{\langle \gamma \rangle_{\rm{for}}^{\star}} \,.
\end{align}

\section{Generalized Powell's relation}
\label{app_powell}

In this appendix, we derive \cref{powell_cond_main} from the main text, we provide an explicit expression of the constant $Z(x_b)$, and we compare our result to the one obtained in \cite{sughiyama_fitness_2019}. To do so, we first derive an alternative form of \cref{powell_cond_main}, for the joint distribution of generation time and newborn size.

\subsection{Generalized Powell's relation with joint probabilities}

We define the steady-state joint distribution $f(\tau,x_b)$ of generation time $\tau$ and newborn size $x_b$ as the ratio of the number of cells born with size $x_b$ and dividing at age $\tau$ at a given snapshot time, to the total number of cells dividing in this snapshot, weighted in either the backward sampling or forward sampling conditioned on survival:
\be
\label{def_f_back_death}
f(\tau,x_b)= \frac{r(\tau,x_b) p(\tau,x_b)}{\int \di \tau' \di x_b' \ r(\tau',x_b') p(\tau',x_b')} \,,
\ee
where the joint distributions of age $a$ and newborn size $x_b$ are the steady-state solutions to \cref{eq_PBE_death_back,eq_for_tim_corr} and read
\begin{align}
	\label{p_back_sol}
	p_{\rm{back}}(a,x_b)=p_{\rm{back}}(0,x_b) \exp \lbk -\Lambda a - \int_{0}^{a} \di a' \ \lp r(a',x_b) + \gamma(a',x_b) \rp \rbk \\
	\label{p_for_sol}
	p^{\star}_{\rm{for}}(a,x_b)=p^{\star}_{\rm{for}}(0,x_b) \exp \lbk -\Gamma a - \int_{0}^{a} \di a' \ \lp r(a',x_b) + \gamma(a',x_b) \rp \rbk \,.
\end{align}
These distributions are independent of the time of the snapshot in steady-state.
Integrating the boundary conditions \cref{eq_PBE_death_back_BC,eq_for_tim_BC_corr_death} over $x_b$ and using the normalization of kernel $\Sigma$, namely $\forall (a',x_b'), \ \int \di x_b \ \Sigma(x_b|x_b',a') =1$, we get that the denominators of \cref{def_f_back_death} are given by:
\begin{align}
	\int \di x_b \ p_{\rm{back}}(0,x_b) &= m \int \di \tau \di x_b' r(\tau,x_b') p_{\rm{back}}(\tau,x_b') \\
	\int \di x_b \ p^{\star}_{\rm{for}}(0,x_b) &= \int \di \tau \di x_b' r(\tau,x_b') p^{\star}_{\rm{for}}(\tau,x_b') \,.
\end{align}
Then, we identify the distribution of newborn size:
\be
\rho^{\rm{nb}}(x_b)=\frac{p(0,x_b)}{\int \di x_b' \ p(0,x_b')} \,,
\ee
for both the forward and backward probabilities.

Finally, combining the results above we obtain
\begin{align}
	\label{res_f_back_death}
	f_{\rm{back}}(\tau,x_b)&= m \rho^{\rm{nb}}_{\rm{back}}(x_b) r(\tau,x_b) \exp \lbk -\Lambda \tau - \int_{0}^{\tau} \di a \ \lp r(a,x_b) + \gamma(a,x_b) \rp \rbk\\
	\label{res_f_for_death}
	f^{\star}_{\rm{for}}(\tau,x_b)&= \rho^{\rm{nb},\star}_{\rm{for}}(x_b) r(\tau,x_b) \exp \lbk -\Gamma \tau - \int_{0}^{\tau} \di a \ \lp r(a,x_b) + \gamma(a,x_b) \rp \rbk \,,
\end{align}
and the generalized Powell's equation reads
\be
\label{powell_joint}
f_{\rm{back}}(\tau,x_b)= m \frac{\rho^{\rm{nb}}_{\rm{back}}(x_b)}{\rho^{\rm{nb},\star}_{\rm{for}}(x_b)}  f^{\star}_{\rm{for}}(\tau,x_b) e^{-(\Lambda-\Gamma) \tau}  \,.
\ee
In the absence of mother-daughter correlations, that is when $\Sigma(x_b|x_b',a') \equiv \hat{\Sigma}(x_b)$, the newborn distributions are unbiased:  $\rho^{\rm{nb}}_{\rm{back}}=\rho^{\rm{nb},\star}_{\rm{for}}=\hat{\Sigma}$, which is a direct consequence of the boundary conditions. Therefore, the fraction in \cref{powell_joint} is canceled, and \cref{powell_joint} can be integrated over $x_b$ to recover Powell's relation in presence of death but without correlations: $f_{\rm{back}}(\tau)= m  f^{\star}_{\rm{for}}(\tau) e^{-(\Lambda-\Gamma) \tau} $.

\subsection{Generalized Powell's relation with conditional probabilities}

We now define the distributions $f(\tau|x_b)$ of generation time conditioned on newborn size $x_b$ as:
\begin{align}
	\label{def_f_cond}
	f(\tau|x_b) &= \frac{r(\tau,x_b) p(\tau,x_b)}{\int \di \tau' \ r(\tau',x_b) p(\tau',x_b)}  \\
	&= \frac{f(\tau,x_b)}{ \rho^{\rm{d}}(x_b) } \,,
\end{align}
where we identified the distribution of newborn sizes at division:
\be
\rho^{\rm{d}}(x_b)=\frac{\int \di \tau' \ r(\tau',x_b) p(\tau',x_b)}{\int \di \tau' \di x_b' \ r(\tau',x_b') p(\tau',x_b')} \,.
\ee

The conditioned distributions are thus given by:
\begin{align}
	\label{res_f_back_cond}
	f_{\rm{back}}(\tau|x_b)&= m \frac{\rho^{\rm{nb}}_{\rm{back}}(x_b)}{\rho^{\rm{d}}_{\rm{back}}(x_b)} r(\tau,x_b) \exp \lbk -\Lambda \tau - \int_{0}^{\tau} \di a \ \lp r(a,x_b) + \gamma(a,x_b) \rp \rbk\\
	\label{res_f_for_cond}
	f^{\star}_{\rm{for}}(\tau|x_b)&= \frac{\rho^{\rm{nb},\star}_{\rm{for}}(x_b)}{\rho^{\rm{d},\star}_{\rm{for}}(x_b)} r(\tau,x_b) \exp \lbk -\Gamma \tau - \int_{0}^{\tau} \di a \ \lp r(a,x_b) + \gamma(a,x_b) \rp \rbk \,,
\end{align}
and finally, Powell's relation on conditional distributions reads:
\be
\label{powell_cond}
f_{\rm{back}}(\tau|x_b)= \frac{m  f^{\star}_{\rm{for}}(\tau|x_b) e^{-(\Lambda-\Gamma) \tau}}{Z(x_b)}  \,,
\ee
with the normalization constant
\be
Z(x_b)=\frac{\rho^{\rm{nb},\star}_{\rm{for}}(x_b)}{\rho^{\rm{nb}}_{\rm{back}}(x_b)} \frac{ \rho^{\rm{d}}_{\rm{back}}(x_b)}{ \rho^{\rm{d},\star}_{\rm{for}}(x_b)} \,.
\ee

\subsection{Comparison with Sughiyama's work}
\label{app_sughi}

In \cite{sughiyama_fitness_2019}, the authors also derived a generalized Powell's relation in the presence of death and mother-daughter correlations. Unlike in our model, they allowed the number of daughter cells at each division to be stochastic and added a probability of death upon division in addition to the death rate between divisions. Moreover, their relation compares the backward distribution in the presence of death to the forward distribution in the absence of death, that we note $f^{\circ}_{\rm{for}}(\tau|x_b)$. In the simple case where there is no death upon division and where the number of daughter cells is fixed to $m$, their result reads
\be
\label{powell_cond_sugi}
f_{\rm{back}}(\tau|x_b)= \frac{m  f^{\circ}_{\rm{for}}(\tau|x_b) e^{-\Lambda \tau - \int_{0}^{\tau} \di a \ \gamma(a,x_b)}}{\tilde{Z}(x_b)}  \,.
\ee

We show here that their relation can be recovered from \cref{powell_cond}. 
In the absence of death, the forward distribution is simply given by
\be
\label{res_f_for_cond_nodeath}
f^{\circ}_{\rm{for}}(\tau|x_b)= r(\tau,x_b) \exp \lbk - \int_{0}^{\tau} \di a \  r(a,x_b) \rbk \,,
\ee
which can also be seen from the normalization of $f^{\circ}_{\rm{for}}$ in \cref{res_f_for_cond}, which imposes $\rho^{\rm{nb},\circ}_{\rm{for}}(x_b)=\rho^{\rm{d},\circ}_{\rm{for}}(x_b)$. Therefore, by comparing \cref{res_f_for_cond,res_f_for_cond_nodeath} we obtain
\be
\label{eq_p_circ_star_pow}
f^{\star}_{\rm{for}}(\tau|x_b) = \frac{\rho^{\rm{nb},\star}_{\rm{for}}(x_b)}{\rho^{\rm{d},\star}_{\rm{for}}(x_b)} f^{\circ}_{\rm{for}}(\tau|x_b) \exp \lbk - \Gamma \tau - \int_{0}^{\tau} \di a \ \gamma(a,x_b) \rbk \,.
\ee
Finally, we plug \cref{eq_p_circ_star_pow} into \cref{powell_cond} to obtain \cref{powell_cond_sugi}, with 
\be
\tilde{Z}(x_b)=Z(x_b) \frac{\rho^{\rm{d},\star}_{\rm{for}}(x_b)}{\rho^{\rm{nb},\star}_{\rm{for}}(x_b)} = \frac{\rho^{\rm{d}}_{\rm{back}}(x_b)}{\rho^{\rm{nb}}_{\rm{back}}(x_b)} \,.
\ee

We believe that our result, \cref{powell_cond}, has the following advantages over Sughiyama's result. First, it compares two distributions evaluated in the same experimental setup, in the presence of death, which makes sense in the light of the definition of fitness proposed in the main text. Second, we obtained explicit expressions for the constants $Z(x_b)$ and $\tilde{Z}(x_b)$, in terms of the distributions of newborn $x_b$ at birth and at division. Third, in \cite{sughiyama_fitness_2019} the authors needed the Direction-Time Hypothesis, which imposes that the kernel is independent of the age at division: $\Sigma(x_b|x_b',a') \equiv \Sigma(x_b|x_b')$. In their article, $x_b$ is not the size at birth, but a general phenotypic or genotypic type, within the multitype age model. Importantly, this assumption implies that their model cannot account for the sizer and adder models, since in these models the kernel should describe the partitioning of volume at division $\Sigma(x_b|x')$, where both age $a'$ and newborn size $x_b'$ are needed to compute mother size $x'$, like detailed in \cref{sec_size_control}. On the other hand, we did not need this assumption.

\section{Adder property concealed by survival}
\label{app_adder} 

In this appendix, we derive the bias between the forward distributions of added volume between birth and division $\Delta_d$ conditioned on birth size $x_b$ in the presence and absence of death, \cref{eq_p_cond_Del_main}.
This distribution
is given by the following change of variable
\be
\label{eq_p_cond_delta_cdv}
f^{\star}_{\rm{for}}(\Delta_d|x_b) \di \Delta_d = f^{\star}_{\rm{for}}(\tau|x_b) \di \tau \,,
\ee
where the distribution of generation time $\tau$ conditioned on newborn size $x_b$ is given by \cref{res_f_for_cond}:
\be
\label{eq_p_cond_tau}
f^{\star}_{\rm{for}}(\tau|x_b)= \frac{\rho^{\rm{nb},\star}_{\rm{for}}(x_b)}{\rho^{\rm{d},\star}_{\rm{for}}(x_b)} \tilde{r}(x(\tau),x(\tau)-x_b) \exp \lbk - \Gamma\tau  -\int_{0}^{\tau} \lbk \tilde{r}(x(t),x(t)-x_b) + \tilde{\gamma}(x(t),x(t)-x_b) \rbk \di t \rbk \,.
\ee
Using \cref{eq_r_adder} and \cref{eq_p_cond_delta_cdv} with $\di \Delta_d / \di \tau = \nu(x(\tau))$, we get
\be
\label{eq_p_cond_Del}
f^{\star}_{\rm{for}}(\Delta_d|x_b)= \frac{\rho^{\rm{nb,\star}}_{\rm{for}}(x_b)}{\rho^{\rm{d},\star}_{\rm{for}}(x_b)} \zeta(\Delta_d) \exp \lbk - \Gamma\tau(x_b,\Delta_d)  -\int_{0}^{\Delta_d} \lbk \zeta(\Delta) + \frac{\tilde{\gamma}(x_b+\Delta,\Delta)}{\nu(x_b+\Delta)} \rbk \di \Delta \rbk \,,
\ee
where we identify the distribution of added volume in the absence of death:
\be
f^{\circ}_{\rm{for}}(\Delta_d)= \zeta(\Delta_d) \exp \lbk - \int_{0}^{\Delta_d} \di \Delta \  \zeta(\Delta) \rbk \,,
\ee
and where we define the normalization constant 
\be
Y(x_b)=\frac{\rho^{\rm{d},\star}_{\rm{for}}(x_b) }{\rho^{\rm{nb,\star}}_{\rm{for}}(x_b)} \,.
\ee
Finally, we recovered \cref{eq_p_cond_Del_main} from the main text. 

To complement this result, let us give several simple examples where the adder property actually remains observable with the forward distribution conditioned on survival. 
First, when the increment of volume $\Delta_d$ between birth and division is fixed, then the distribution of added volume is a Dirac delta function, independent of the size at birth, for every distribution.
Second, in the case of a uniform death/dilution rate $\tilde{\gamma}(x,\Delta) \equiv \gamma$, we showed in \cref{eq_link_Gam_gam} that $\Gamma=-\gamma$.  Therefore, the two terms in the exponential bias cancel out, as $\int_{0}^{\Delta_d} \di \Delta  \ \nu(x_b+\Delta)^{-1} = \tau(x_b,\Delta_d)$. Once again, we find that a constant death rate does not affect quantitatively the observables. 
Third, the exponential bias becomes independent of the newborn volume when both of its terms are independent of $x_b$. This is the case when 
(i) cells grow linearly with a rate $\nu(x)=\nu_1$ such that: $\tau(x_b,\Delta_d) \equiv \tau(\Delta_d) =\Delta_d/\nu_1$, and (ii) the death rate obeys a similar separation of variables as the division rate, namely $\tilde{\gamma}(x,\Delta) \equiv \nu(x)\xi(\Delta)$ with $\xi(\Delta)$ the death rate per unit volume. 
In the second and third cases, note that the independence of the exponential bias from $x_b$ implies that the factor $Y(x_b)$ is also independent of $x_b$. This is because the normalization of the distribution: $ \forall x_b, \ \int_0^{\infty} \di \Delta_d \ f^{\star}_{\rm{for}}(\Delta_d|x_b) =1$, imposes $Y(x_b) \equiv Y$.

Finally, let us mention that it has also been proposed in \cite{amir_cell_2014} to test the adder property on the basis of the value of the Pearson correlation coefficient between mother and daughter properties (typically size at birth), instead of the independence of the lineage distribution of added volume from the newborn volume. This alternative test has for example been used in \cite{taheri-araghi_cell-size_2015} to complement the analysis, leading to the same conclusion that data were best accounted for by the adder model. It remains open to investigate the effect of the survivor bias on the Pearson correlation coefficient.

\end{document}